\documentclass[10pt,journal,compsoc]{IEEEtran}

\ifCLASSOPTIONcompsoc
\usepackage[nocompress]{cite}
\else
\usepackage{cite}
\fi

\ifCLASSINFOpdf
\else
\fi

\usepackage[english]{babel}

\usepackage{booktabs} % For formal tables
\usepackage[numbers]{natbib}

\usepackage{array}
\newcolumntype{P}[1]{>{\centering\arraybackslash}p{#1}}
\usepackage{romannum}
\usepackage{balance}
 \usepackage{graphicx,multirow}
\usepackage{multirow}
\usepackage{graphicx}
\usepackage{tabularx}
\usepackage[table,xcdraw]{xcolor}
\usepackage{xcolor, colortbl}
\usepackage{xspace}
\usepackage[flushleft]{threeparttable}
\usepackage{subcaption}
\usepackage{listings,xcolor}
\usepackage[hyphens]{url}

\definecolor{grayboxcolor}{HTML}{f2f2f2}
\newcommand{\conclusion}[1]{%
	\begin{center}\noindent\thicklines\setlength{\fboxsep}{8pt}\fcolorbox{black}{grayboxcolor}{\begin{minipage}{3.3in}\textit{\textbf{#1}}\end{minipage}}\end{center}} 

\definecolor{awesome}{rgb}{1.0, 0.13, 0.32}

\definecolor{blue-violet}{rgb}{0.54, 0.17, 0.89}

\begin{document}
\pagestyle{headings}
\pagenumbering{arabic}

% Intents and Entities Styling Command
\newcommand{\Style}{\textit}% 

% Repository Intents Commands
\newcommand{\CountCommitsByDates}{\Style{CountCommitsByDate}\xspace}% 
\newcommand{\FileCommits}{\Style{FileCommits}\xspace}% 
\newcommand{\OverloadedDev}{\Style{OverloadedDev}\xspace}% 
\newcommand{\CommitsByDate}{\Style{CommitsByDate}\xspace}% 
\newcommand{\ExperiencedDevFixBugs}{\Style{ExperiencedDevFixBugs}\xspace}% 
\newcommand{\BuggyFiles}{\Style{BuggyFiles}\xspace}% 
\newcommand{\BuggyCommitsByDate}{\Style{BuggyCommitsByDate}\xspace}% 
\newcommand{\FixCommit}{\Style{FixCommit}\xspace}%
\newcommand{\BuggyCommit}{\Style{BuggyCommit}\xspace}
\newcommand{\BuggyFixCommits}{\Style{BuggyFixCommit}\xspace}

% SOF Intents Commands
\newcommand{\UsingMethodImproperly}{\Style{UsingMethodImproperly}\xspace}% 
\newcommand{\LookingForCodeSample}{\Style{LookingForCodeSample}\xspace}% 
\newcommand{\FacingError}{\Style{FacingError}\xspace}%
\newcommand{\PassingData}{\Style{PassingData}\xspace}% 
\newcommand{\LookingForBestPractice}{\Style{LookingForBestPractice}\xspace}% 

% Repository Entities Commands
\newcommand{\FileName}{\Style{FileName}\xspace}% 
\newcommand{\JiraTicket}{\Style{JiraTicket}\xspace}% 
\newcommand{\DateTime}{\Style{DateTime}\xspace}%
\newcommand{\CommitHash}{\Style{CommitHash}\xspace}% 

% SOF Entities Commands
\newcommand{\ProgLanguage}{\Style{ProgLanguage}\xspace}% 
\newcommand{\Framework}{\Style{Framework}\xspace}% 
\newcommand{\Standards}{\Style{Standards}\xspace}%
\newcommand{\API}{\Style{API}\xspace}% 
\newcommand{\Platform}{\Style{Platform}\xspace}% 

% Additional Intents Commands
\newcommand{\FacingException}{\Style{FacingException}\xspace}% 
\newcommand{\GetMaintainers}{\Style{GetMaintainers}\xspace}% 
\newcommand{\GetFixingCommits}{\Style{GetFixingCommits}\xspace}%

\title{A Comparison of Natural Language Understanding Platforms for Chatbots in Software Engineering}

\author{Ahmad~Abdellatif, Khaled~Badran, Diego~Elias~Costa, and Emad~Shihab,~\IEEEmembership{Senior Member,~IEEE}

\IEEEcompsocitemizethanks{\IEEEcompsocthanksitem Ahmad Abdellatif, Khaled Badran, Diego Elias Costa, and Emad Shihab are with the Data-driven Analysis of Software (DAS) Lab at the Department of Computer Science and Software Engineering, Concordia University, Montr\'{e}al, Canada.
	
	E-mail: {a\_bdella, k\_badran, d\_damasc, eshihab}@encs.concordia.ca}}

%\markboth{IEEE TRANSACTIONS ON SOFTWARE ENGINEERING, February~2021}%
%{Shell \MakeLowercase{\textit{et al.}}: Bare Demo of IEEEtran.cls for Computer Society Journals}

\IEEEtitleabstractindextext{
\begin{abstract}
Chatbots are envisioned to dramatically change the future of Software Engineering, allowing practitioners to chat and inquire about their software projects and interact with different services using natural language.
At the heart of every chatbot is a Natural Language Understanding (NLU) component that enables the chatbot to understand natural language input. 
Recently, many NLU platforms were provided to serve as an off-the-shelf NLU component for chatbots, 
however, selecting the best NLU for Software Engineering chatbots remains an open challenge.

Therefore, in this paper, we evaluate four of the most commonly used NLUs, namely IBM Watson, Google Dialogflow, Rasa, and Microsoft LUIS to shed light on which NLU should be used in Software Engineering based chatbots. Specifically, we examine the NLUs' performance in classifying intents, confidence scores stability, and extracting entities. To evaluate the NLUs, we use two datasets that reflect two common tasks performed by Software Engineering practitioners, 1) the task of chatting with the chatbot to ask questions about software repositories 2) the task of asking development questions on Q\&A forums (e.g., Stack Overflow). According to our findings, IBM Watson is the best performing NLU when considering the three aspects (intents classification, confidence scores, and entity extraction). However, the results from each individual aspect show that, in intents classification, IBM Watson performs the best with an F1-measure$>$84\%, but in confidence scores, Rasa comes on top with a median confidence score higher than 0.91. Our results also show that all NLUs, except for Dialogflow, generally provide trustable confidence scores. For entity extraction, Microsoft LUIS and IBM Watson outperform other NLUs in the two SE tasks. Our results provide guidance to software engineering practitioners when deciding which NLU to use in their chatbots.

\end{abstract}
}

\maketitle
\IEEEdisplaynontitleabstractindextext
\IEEEpeerreviewmaketitle

\section{Introduction}
\label{sec:intro}
\IEEEPARstart{S}{oftware} chatbots are increasingly used in the Software Engineering (SE) domain since they allow users to interact with  platforms using natural language, automate tedious tasks, and save time/effort \cite{Storey_2016FSE}. This increase in attention is clearly visible in the increase of number of bots related publications \cite{Xu_2017ASE,Abdellatif2019EMSE,Dominic_2020BotSE,Abdellatif_MSR2020,Lin_2020BotSE}, conferences \cite{ChatbotConference}, and workshops \cite{BotSE2019}. A recent study showed that one in every four OSS projects (26\%) on GitHub are using software bots for different tasks \cite{Wessel_2018CSCW}. This is supported by the fact that bots help developers perform their daily tasks more efficiently\cite{Storey_2016FSE}, such as deploy builds \cite{DeployBot_link}, update dependencies \cite{Dependabot_link}, and even generate fixing patches~\cite{Urli_2018ICSE-SEIP}.

At the core of all chatbots lie the Natural Language Understanding platforms---referred hereafter simply as NLU.  NLUs are essential for the chatbot's ability to understand and act on the user's input \cite{8554770,braun_SIGDIAL2017}. 
The NLU uses machine-learning and natural language processing (NLP) techniques to extract structured information (the intent of the user's query and related entities) from unstructured user's input (textual information).
As developing an NLU from scratch is very difficult because it requires NLP expertise, chatbot developers resort to a handful of widely-used NLUs that they leverage in their chatbots~\cite{Abdellatif2019EMSE,8500065,marbotSe78:online,Murgia_2016CHI,Toxtli_2018CHI}.

As a consequence of the diversity of widely-used NLUs, developers are faced with selecting the best NLU for their particular domain. This is a non-trivial task and has been discussed heavily in prior work (especially since NLUs vary in performance in different contexts)~\cite{braun_SIGDIAL2017,canonico_CC2018,gregori_2017GIT}.
For instance, in the context of the weather domain, \citet{canonico_CC2018} showed that IBM Watson outperformed other NLUs, while \citet{gregori_2017GIT} evaluated NLUs using frequently asked questions by university students and found that Dialogflow performed best.
In fact, there is no shortage of discussions on Stack Overflow about the best NLU to use in chatbot implementation~\cite{Naturall58:online,artifici16:online,nlpBuild16:online} as choosing an unsuitable platform for a particular domain deeply impacts the user satisfaction with the chatbot~\cite{Lebeuf_2018IST,NegBotImpact_Report,TheimpactofBot_Link}.

An important domain that lacks any investigation over different NLUs' performance is SE.
Software Engineering is a specialized domain with very specific terminology that is used in a particular way. 
For example, in the SE domain, the word `ticket' refers to an issue in a bug tracking system (e.g., Jira), while in other domains it is related to a movie (e.g., TicketMaster bot) or flight ticket. Moreover, there is no consensus amongst SE chatbot developers on the best NLU to use for the SE domain. For instance, TaskBot~\cite{Toxtli_2018CHI} uses Microsoft Language Understanding Intelligent Service (LUIS)~\cite{LUISLang29:online} to help practitioners manage their tasks. MSRBot~\cite{Abdellatif2019EMSE} uses Google Dialogflow NLU to answer questions related to the software repositories. MSABot~\cite{Lin_2020BotSE} leverages Rasa NLU to assist practitioners in developing and maintaining microservices. Given that no study has investigated which NLU performs best in the SE domain, chatbot developers can not make an informed decision on which NLU to use when developing SE-based chatbots.

Hence, in this paper, we provide the first study to assess the performance of widely-used NLUs to support SE tasks. We evaluate NLUs on queries related to two important SE tasks: 1) Repository: Exploring projects' repository data (e.g.,``What is the most buggy file in my repository?''), and 2) Stack Overflow: Technical questions developers frequently ask and answer from Q\&A websites (e.g., ``How to convert XElement object into a dataset or datatable?'').

Using the two SE tasks, we evaluate four widely-used NLUs: IBM Watson~\cite{IBMWatso77:online}, Google Dialogflow~\cite{Dialogfl49:online}, Rasa~\cite{RasaOpen48:online}, and Microsoft LUIS~\cite{LUISLang29:online} under three aspects:
1)  the NLUs' performance in correctly identifying the purpose of the user query (i.e., intents classification);
2)  the confidence yielded by the NLUs when correctly classifying and misclassifying queries (i.e., confidence score); and
3) the performance of the NLUs in identifying the correct subjects from queries (i.e., entity extraction).

Our results show that, overall (considering NLUs' performance in intents classification, confidence score, and entity extraction), IBM Watson is the best performing NLU for the studied SE tasks. However, the findings from evaluating the NLUs on individual aspects show that the best performing NLU can vary. IBM Watson outperforms other NLUs when classifying intents for both tasks (F1-measure $>$ 84\%). Also, we find that all NLUs (except for Dialogflow in one task) report high confidence scores for correctly classified intents. Moreover, Rasa proves to be the most trustable NLU with a median confidence score $>$ 0.91. When extracting entities from SE tasks, no single NLU outperforms the others in both tasks. LUIS performs the best in extracting entities from the Repository task (F1-measure 93.7\%), while IBM Watson comes on top in the Stack Overflow task (F1-measure 68.5\%).

Given that each NLU has its own strengths in the different SE tasks (i.e., performs best in intent classification vs. entity extraction), we provide an in-depth analysis of the performance of the different NLU's features, {which are the list feature, where the NLU extracts entities using an exact match from a list of synonyms; and the prediction feature, where the NLU predicts entities that it might not have been trained on before.} Also, we analyze the characteristics of the intents in each task to better understand the intents that tend to be harder to classify by all of the evaluated NLUs.  

The paper makes the following contributions:
\begin{itemize}
\item[$\bullet$] To the best of our knowledge, this is the first work to evaluate NLUs on two representative tasks (i.e., software repositories data and Stack Overflow posts) from the SE domain.
\item[$\bullet$] We evaluate the NLUs using different features for extracting entities (i.e., list and prediction features).
\item[$\bullet$] We explore the impact of selecting different confidence score thresholds on the NLUs' intent classification performance.
\item[$\bullet$] We provide a set of actionable recommendations, based on our findings and  experience in conducting this study, for chatbot practitioners to improve their NLU's performance.
\item[$\bullet$] We make our labelled dataset publicly available to enable replication and help advance future research in the field~\cite{NLUsCompScript_Link}.
\end{itemize}

The rest of the paper is organized as follows: Section~\ref{sec:background} provides an overview about chatbots and explains related concepts used throughout this paper. Section~\ref{sec:casesetup} describes the case study setup used to evaluate the performance of the NLUs. We report the evaluation results in Section~\ref{sec:results}. Section~\ref{sec:discussion} discusses our findings and provides a set of recommendations to achieve better classifications results.
Section~\ref{sec:relatedwork} presents the related work to our study. Section~\ref{sec:threats} discusses the threats to validity, and section~\ref{sec:conclusions} concludes the paper.

\section{Background}
\label{sec:background}

Before diving into the NLUs' evaluation, we explain in this section the chatbot-related terminology used throughout the paper. We also present an overview of how chatbots and NLUs work together to perform certain actions.

\subsection{Definitions}

Software chatbots are the conduit between their users and automated services \cite{Lebeuf_2018IST}. Through natural language, users ask the chatbot to perform specific tasks or inquire about a piece of information. 
Internally, a chatbot then uses the NLU to analyze the posed query and act on the users' request. 
The main goal of an NLU is to extract structured data from unstructured language input. 
In particular, it extracts intents and entities from users' queries: intents represent the user intention/purpose of the question, while entities represent important pieces of information in the query. 
For example, take a chatbot like the MSRBot~\cite{Abdellatif2019EMSE}, that replies to user queries about software repositories. 
In the query ``How many commits happened in the last month of the project?", the intent is to know the number of commits that happened in a specific period (\CountCommitsByDates), and the entity `last month' of type \DateTime determines the parameter for the query.
The chatbot uses both the intent and entities to perform the action that answers the user's question. In this example, the chatbot searches in the repository for the number of commits issued in the last month.

Most NLUs come with a set of built-in entities (e.g., currencies and date-time), which are pre-trained on general domain queries. 
To use an NLU on a specialized domain, developers should define a set of custom intents and entities. 
For each custom intent, the NLU needs to be trained on a set of queries that represents different ways a user could express that intent. 
Again, taking the former example, ``How many commits happened in the last month?'', this query can be asked in multiple different ways. 
For instance, ``show me the number of commits between 1-03-2020 and 31-03-2020'' is an example of a query with the same semantics but different syntax.
Both queries can and should be used to train the NLU on how to identify the \CountCommitsByDates intent.
Similarly to custom intents, NLUs need to be trained to recognize custom entities.
To do that, developers label the entity types and their values in the queries. 
For example, in the following query ``what is the fixing commit for bug HHH-8501?'', the entity `HHH-8501' is labelled as a \JiraTicket type.

The misclassification of intents and entities negatively impacts the user experience, although each in its own way.
When an NLU misclassifies an intent, the chatbot fails to understand the query in a fundamental manner, leading the chatbot to reply to a different query or performing the wrong task.
Misclassifying entities, on the other hand, causes the chatbot to reply about a wrong piece of information. 
For example, in the query ``How to convert xml to json file in java'' there are three entities: `XML', `Json' and `Java'. If the NLU fails to extract the `Java' entity, the chatbot loses the context of the question and might reply with an answer for converting XML to Json with code example from any other programming language (e.g., Python).

The last piece in the picture is the confidence score, which represents how confident the NLU is in classifying the intent \cite{RasaConfidence_link, LUISConfidence_link, DialogflowConfidence_link, WatsonConfidence_link}. The confidence score is given on a scale from 0 (i.e., not confident) to 1 (i.e., fully confident), which corresponds to the classified intent by the NLU. Chatbot developers use the confidence score to choose their next action, either by answering the user's question/request or triggering a fallback intent. The fallback intent is a response issued by the chatbot to give the user a chance to rephrase or clarify their initial query. 
Typically, the fallback intent is triggered when the returned confidence score is lower than a certain threshold. 
Choosing a suitable threshold for a chatbot is not an easy task, as a low value would make a chatbot answer to unclear questions more often (too confident), and a high threshold would trigger the fallback intent too often (insecure chatbot), annoying the user by asking it to rephrase the question frequently.

In our study, we want to investigate the NLUs' performance with regards to intents classification, confidence score, and entity extractions. 
All three aspects are critical to ensure that chatbots return correct and complete responses to the user.

\begin{figure*}[t!]
	\centering
	\includegraphics[scale=0.58]{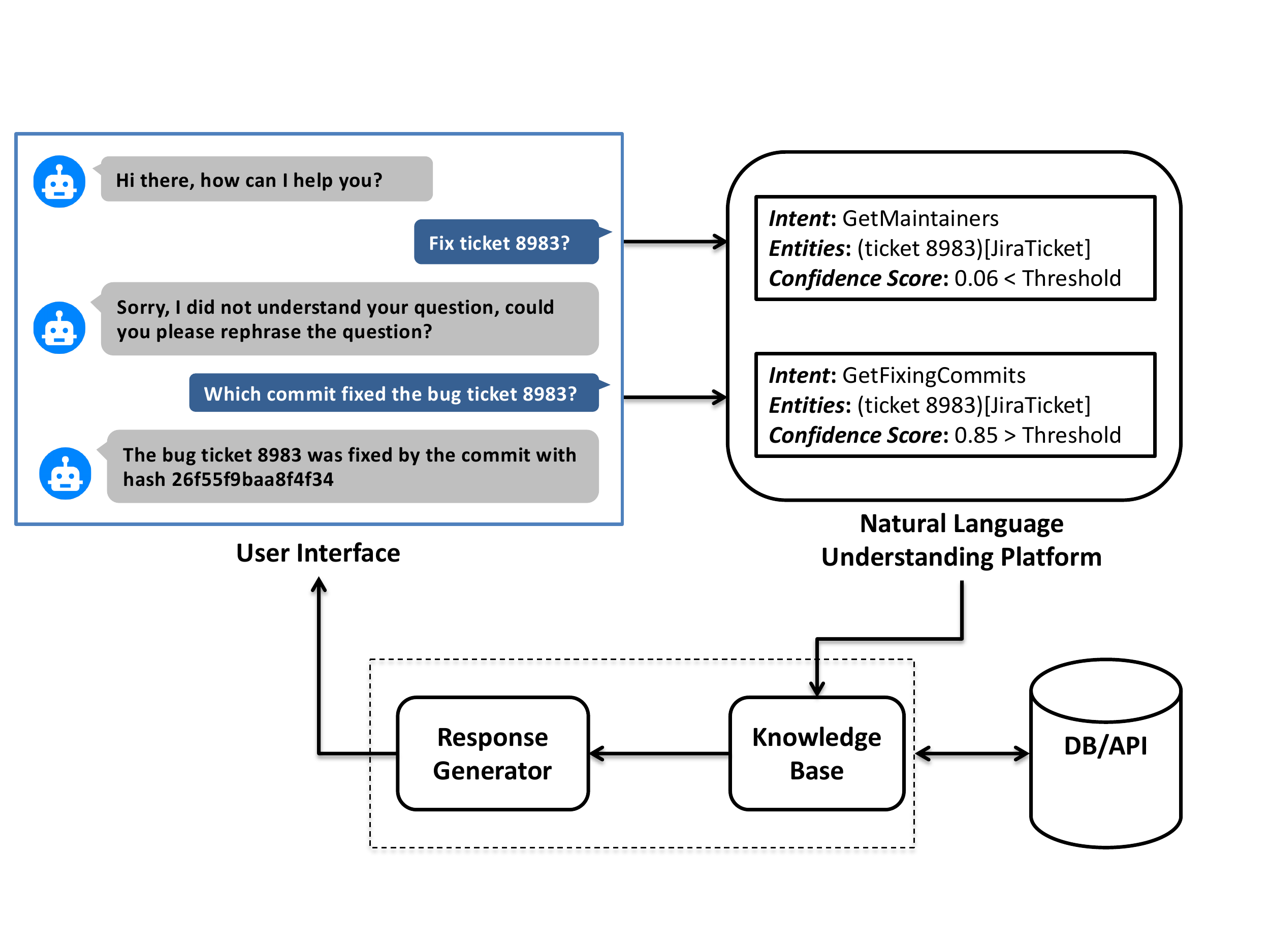}
	\vspace{-0.3in}
	\caption{An overview of user-chatbot interaction}
	\label{fig:workingExample}
\end{figure*}

\subsection{Explanatory Example}

To demonstrate how chatbots utilize NLUs to answer a user's query, we showcase an example of a user asking a repository related question to a chatbot as shown in Figure~\ref{fig:workingExample}. In this example, we use a simplified architecture of the chatbot~\cite{Abdellatif2019EMSE} for illustration purposes. 
The NLU is trained on the queries (intents) related to mining software repositories and is trained to extract repository entities from users' questions, such as a \JiraTicket (e.g., HHH-8593).
In this example, after the costumary greeting from the chatbot, the user asks the chatbot ``Fix ticket 8983?'' which is forwarded to the NLU where it classifies the user's question as having a \GetMaintainers intent with a confidence score of 0.06. 
The low confidence score (lower than a predetermined threshold) triggers the fallback intent,
thus, the chatbot asks the user to rephrase the question in a more understandable way (i.e., ``Sorry, I did not understand your question, could you please rephrase the question?''). 
After the user rephrases the question ``Which commit fixed the bug ticket 8983?'', the NLU extracts the entity `ticket 8983' of type \JiraTicket and classifies the intent of the query as \GetFixingCommits with a confidence score of 0.85. Finally, the chatbot performs the necessary action, querying the database to answer the posed question (``The bug ticket 8983 was fixed by the commit with hash 26f55f9baa8f4f34").

\section{Case Study Setup}
\label{sec:casesetup}
Since the main goal of this paper is to evaluate the performance of different NLUs using SE tasks, we need to select the candidate NLUs that we want to examine and the SE tasks' data corpus to train and test those NLUs. In this section, we detail our selection of the NLUs, SE tasks used in the evaluation, and our experiment design.

\subsection{Evaluated NLUs}

There exists several widely-used NLUs that are easily integrated with third-party applications. To make our study comprehensive, we choose to examine the performance of four NLUs, namely IBM Watson, Dialogflow, Rasa, and LUIS. 
We select these NLUs since they are popular and widely used by both researchers and practitioners~\cite{8500065, Toxtli_2018CHI}, and have been studied by prior NLU comparison work in other domains~\cite{braun_SIGDIAL2017, gregori_2017GIT, koetter_2018ICART}.
Moreover, all selected NLUs can be trained by importing the data through their user interface or API calls, which facilitates the training process. 
In the following, we provide a description of those NLUs.

\begin{itemize}
	
	\item \textbf{Watson Conversation (IBM Watson):} An NLU provided by IBM \cite{IBMWatso77:online}. IBM Watson has prebuilt models for different domains (e.g. banking) and a visual dialog editor to simplify building the dialog by non-programmers.	
		
	\item \textbf{Dialogflow:} An NLU developed by Google \cite{Dialogfl49:online}. Dialogflow supports more than 20 spoken languages and can be integrated with many chatting platforms such as Slack~\cite{Dialogfl49:online}.
	
	\item \textbf{Rasa:} The only open-source NLU in our study, owned by Rasa Technologies \cite{RasaOpen48:online}. Rasa allows developers to configure, deploy, and run the NLU on local servers. Thus, increasing the processing speed by saving the network time compared to cloud-based platforms. 
In our evaluation, we use Rasa-nlu v0.14, which was the latest version when conducting the experiment.

	\item\textbf{Language Understanding Intelligent Service (LUIS):} An NLU cloud platform from Microsoft \cite{LUISLang29:online}. LUIS has several prebuilt domains such as music and weather, and supports five programming languages: C\#, Go, Java, Node.js, and Python.

\end{itemize}

\subsection{SE Tasks and Data Corpora}
\label{subsec:datasets}

\begin{table*}[tbh]
	\centering
	\caption{Intents distribution in the Repository task.}
	\label{table:chatbotIntent}
	\begin{tabularx}{\linewidth}{l X r r r}
		\toprule
		\textbf{Intent} & \textbf{Definition} & \textbf{Train (\%)} & \textbf{Test (\%)} & \textbf{Total (\%)} \\ 
		\midrule
		\textbf{BuggyCommitsByDate} & Present the buggy commit(s) which happened during a specific time period.   & 66 (23.8) & 13 (10.7) & 79 (19.6) \\
		\textbf{BuggyCommit} & Identify the bugs that are introduced because of certain commits.   & 52 (18.8) & 9 (7.4) & 61 (15.3) \\
		\textbf{BuggyFiles} & Determine the most buggy files in the repository to refactor them.   & 37 (13.4) & 13 (10.7) & 50 (12.6) \\
		\textbf{FixCommit} & Identify the commit(s) which fix a specific bug.   & 31 (11.2) & 11 (9.0) & 42 (10.6) \\
		\textbf{BuggyFixCommits} & Identify the fixing commits that introduce bugs at a particular time   & 32 (11.6) & 7 (5.8) & 39 (9.8) \\ 			
		\textbf{CountCommitsByDates} & Identify the number of commits that were pushed during a specific time period.  & 11 (3.9) & 21 (17.4) & 32 (8.0) \\
		\textbf{ExperiencedDevFixBugs} & Identify the developer(s) who have experience in fixing bugs related to specific file.  & 15 (5.4) & 14 (11.6) & 29 (7.3) \\
		\textbf{OverloadedDev} & Determine the overloaded developer(s) with the highest number of unresolved bugs.   & 15 (5.4) & 9 (7.4) & 24 (6.0) \\
		\textbf{FileCommits} & View details about the changes that are occurred on on a file.   & 10 (3.6) & 12 (10.0) & 22 (5.5) \\
		\textbf{CommitsByDate} & Present the commit information (e.g., commit message) at a specific time.  & 8 (2.9) & 12 (10.0) & 20 (5.0) \\
		\bottomrule
	\end{tabularx}
\end{table*}

To evaluate the performance of the NLUs in the Repository and Stack Overflow tasks, we select two representative data corpora, one for each task 1) \textit{Repository corpus} \cite{Abdellatif2019EMSE} used for the Repository task and includes questions posed to a chatbot by practitioners looking for information related to their projects' software repositories 2) \textit{Stack Overflow corpus} \cite{Ye_2016Saner} used for the Stack Overflow task and contains a set of posts from Stack Overflow discussion threads. 
Our selection of these two tasks was motivated by two main reasons: 
Firstly, both tasks reflect realistic situations, as in both, developers are asking questions about issues they face or to get more information about their projects (e.g., fixing commit for a bug). 
In fact, the Repository task also covers questions that are commonly asked by project managers to grasp the state of the project repository. 
Hence, both tasks make our results more generalizable to the chatbots practitioners in the SE domain. 
Secondly, using two tasks in our evaluation gives us better insights on how each NLU performs in different sub-contexts of SE. The Stack Overflow task uses a corpus that has a diverse set of posts from the top 8 tags in Stack Overflow, the most popular Q\&A website in the developers community \cite{Mamykina_2011CHI, Abdalkareem_2017IEEESoftware}. 
On the other hand, the Repository task contains project specific information (i.e., ``who touched file x?'') rather than general programming questions.
\\

\begin{table}[tb]
	\centering
	\caption{Entities distribution in the Repository task.}
	\label{table:chatbotEntity}
	\begin{tabularx}{\linewidth}{lXrr}                       
		\toprule
		\multicolumn{1}{c}{\textbf{Entity Type}} & \textbf{Definition}  & \textbf{Train} & \textbf{Test} \\
		\midrule
		\textbf{FileName} & Name of the file (e.g., Transaction.java). & 35,007 & 26 \\
		\textbf{JiraTicket} & Ticket ID number (e.g., KAFKA-3612). & 21,012 & 11 \\
		\textbf{DateTime} & Specific/period data (e.g., during July of 2019). & 117 & 52 \\
		\textbf{CommitHash} & Hash of a certain commit. & 15,303 & 10 \\
		\bottomrule
	\end{tabularx}
\end{table}

\noindent\textbf{Repository Corpus.} 
This corpus was originally used to train and test the MSRBot~\cite{Abdellatif2019EMSE}, a chatbot tailored for answering users' questions on project repository data. This corpus contains a set of 398 queries, with 10 different intents, as shown in Table~\ref{table:chatbotIntent}. 
Each intent contains a set of different questions with the same semantic, indicating the different ways a user could ask the same query. Those intents have questions related to code repository (e.g., ``List me the changes done in ClassA.java'' from the \FileCommits intent), issue tracker (e.g., ``Who has the most bug assignments?'' from the \OverloadedDev intent), or a combination of both code and issue tracker (e.g., ``Which commits fixed HHH-10956?'' a query from the \FixCommit intent).
The corpus includes explicitly defined training and test sets that are labelled by the developers of the MSRBot. %\ahmad{Do we need to mention that the MSRBot operated on Hibernate and Kafka?}
The training set includes the questions used to train the chatbot, acquired and curated by three developers involved in the project, while the test set is composed of questions posed by 12 software developers to test the MSRBot on a specified set of 10 different tasks. 
We use the training and test set as defined by the MSRBot authors in this experiment.

Each entity in the Repository corpus contains a so-called list feature~\cite{ListEntityDf_link,ListEntityLuis_link,ListEntityRasa_link,ListEntityIBMWatson_link}, a list of equivalent synonyms used by the NLU to recognize entities. 
With the list feature, the NLUs are limited to extract the exact match of the entities, but can use the specified synonyms in the extraction process. 
For instance, the entity `HHH-7325' of type \JiraTicket has a list of synonyms containing `bug7325', `issue7325', and `HHH7325', hence, any of these words can be used by the NLU to recognize the entity `HHH-7325'.
The Repository corpus contains four entity types using the list feature as shown in Table \ref{table:chatbotEntity}, covering the main artifacts in a repository, such as files (\FileName), commits (\CommitHash), and tickets from the JIRA bug-tracker (\JiraTicket).
\\
\begin{table}[tb]
	\centering
	\caption{List of the Stack Overflow task entities.}% 
	\label{sof_entities}
		\begin{tabularx}{\linewidth}{lXr}
		\toprule
		\textbf{Entity} & \textbf{Definition} & \textbf{Total} \\ 
		\midrule
		\textbf{ProgLanguage}  & Types of programming languages (e.g., Java) \cite{Ye_2016Saner}. & 96 \\
		\textbf{Framework} & Tools/frameworks that developers use (e.g., Maven) \cite{Ye_2016Saner}. & 85 \\
		\textbf{Standards}  & ``Refers to data formats (e.g., JSON), design patterns (e.g.,Abstract Factory), protocols (e.g., HTTP), technology acronyms (e.g., Ajax)'' \cite{Ye_2016Saner}. & 20 \\
		\textbf{API} & An API of a library (e.g., ArrayList) \cite{Ye_2016Saner}.& 67 \\
		\textbf{Platform}  & Software/Hardware platforms (e.g., IOS) \cite{Ye_2016Saner}. & 13 \\ 
		\bottomrule
	\end{tabularx}
\end{table}

\begin{table*}[tbh]
	\caption{List of the Stack Overflow task intents.}
	\centering
	\label{sof_intents}
	\begin{tabularx}{\linewidth}{lXr}
		\toprule
		\multicolumn{1}{c}{\textbf{Intent}} & \textbf{Description} & \textbf{Total (\%)} \\ 
		\midrule
		\textbf{LookingForCodeSample} & Looking for information related to implementation. This includes looking for code snippets, the functionality of a method, or information specific to the user's needs.& 132 (61.3\%) \\
		\textbf{UsingMethodImproperly}  & A method or a framework is being used improperly causing an unexpected or unwanted behaviour of the program in hand. This can be related to code bugs or to performance issues.& 51 (23.7\%) \\
		\textbf{LookingForBestPractice} & Looking for the recommended (best) practice, approach or solution for a problem. & 12 (5.6\%) \\ 
		\textbf{FacingError} &  Facing an error or a failure in a program, mostly in the form of an error message or a build failure.  & 10 (4.7\%)\\
		\textbf{PassingData} &  Passing data between different frameworks or method calls.& 10 (4.7\%) \\
		\bottomrule
	\end{tabularx}
\end{table*}
\noindent\textbf{Stack Overflow Corpus.} Stack Overflow is a popular Q\&A website among developers and plays an important role in the software development life cycle \cite{Abdalkareem_2017IEEESoftware}. 
Given its importance, data from Stack Overflow has already been used in prior work to train chatbots \cite{Xu_2017ASE}. 
The titles of the questions in Stack Overflow represent a request for information by software practitioners 
(e.g., ``How to create an JS object from scratch using a HTML button?'').
Recently, \citet{Ye_2016Saner} developed a machine learning approach to extract software entities (e.g., the programming language name) from Q\&A websites like Stack Overflow, and manually labelled entities from 297 Stack Overflow posts as shown in Table~\ref{sof_entities}. 
We use the same corpus~\cite{Ye_2016Saner} and extract the title of each post with its labelled entities. 
Unlike the Repository corpus, entities in the Stack Overflow corpus do not have a list feature, (list of entity synonyms), and need to be predicted by the NLUs (i.e., prediction feature). In other words, we train the NLUs only on the entities included in the training set. This allows us to evaluate the NLUs' ability to extract entities that they have not been trained on before, which emulates real-life scenarios where it is difficult for the practitioners to train the NLUs on all SE entities.

While the original Stack Overflow corpus contains manually labelled entities, it lacks the intent behind the posed questions. 
Hence, we need to manually label the intents of the queries before we can use the corpus for the Stack Overflow task.
To achieve that, the first author used thematic analysis \cite{Cruzes_2011ESEM} to categorize those queries (titles) according to their intents. Thematic analysis~\cite{braun2006ThematicAnalysis} is a technique to extract common themes within a corpus via inspection, which was done manually in our case.
This method is frequently used in qualitative research to identify patterns within collected data and has been used in prior work in the SE domain \cite{Lenberg:2015:HFR:2819321.2819329, Munir2016}.
Initially, the first author categorized the queries into 19 intents. Then, we merged the categories that have a small number of examples (less than 10) but have a very similar rationale. 
This is because some NLUs recommend a training set of at least 10 queries for each intent \cite{Frequent42:online}. 
For example, the queries of \FacingException and \FacingError intents are quite similar in their goal as developers are looking for a solution to fix the crash and error. 

To validate the manually extracted intents, we asked two additional annotators - the second author and one other Ph.D. student - to independently label the queries using the extracted intents. 
For each question, we asked the annotators to evaluate whether the query has a clear intent or not using the multiple choice (`Yes', `May be', and `No'). 
If the annotators answer the previous question with `Yes' or `May be', then they classify the title using one of the defined intents shown in Table \ref{sof_intents}. 
After both annotators finished the labelling process, we merged the labelled queries into one set to be used in the NLUs evaluation.
We then use the Cohen's Kappa coefficient to evaluate the level of agreement between the two authors~\cite{Cohen_kappa}. The Cohen's Kappa coefficient is a well-known statistic that evaluates the inter-rater agreement level for categorical scales. The resulting coefficient is a scale that ranges between -1.0 and +1.0, where a negative value means poorer than chance agreement, zero indicates agreement by chance, and a positive value indicates better than chance agreement. We find that the level of agreement between the two annotators on the occurrence of intent (i.e., whether a title has an intent or not) is +0.74, and the agreement on the intents classification is +0.71. Both agreements are considered to be substantial inter-rater agreements~\cite{kappa_agreement_level}.
All three annotators 
discussed the disagreements and voted to the best fitting intent.
After the merge, we discarded queries with unclear intent (total of 82), such as ``JConsole Web Application''. 
The final set includes 215 queries \cite{NLUsCompScript_Link}, Tables \ref{sof_entities} and \ref{sof_intents} show the number of entities and intents included in our evaluation, respectively.

Our manual classification led to the creation of 5 intents shown in Table \ref{sof_intents} with their definitions. The intents in the Stack Overflow Corpus represent different types of questions posted on Stack Overflow. For example, the \FacingError intent contains questions asking for a solution for an encountered error:  ``PHP mysqli query returns empty error message''~\cite{diePHPmy88:online}. 
Table~\ref{sof_entities} shows entities used in the Stack Overflow corpus that are very specific to the SE domain. 
For example, \ProgLanguage entity type has a set of different programming languages such as Python and JavaScript.  
Such entities can be used by code-centric chatbots~\cite{Wyrich_2019BotSE,Lin_2020BotSE}, such as chatbots that extract files that call a particular method in the code~\cite{Lin_2020BotSE}.

\subsection{Performance Evaluation of NLUs}

We use the corpora from the Repository and Stack Overflow tasks to train IBM Watson, Dialogflow, Rasa, and LUIS and evaluate their performance. 
As each task has specific characteristics, we detail in the following 
how we train and test the NLUs for each task.

To evaluate the NLUs on the Repository task, we use the same training set from the Repository corpus, which includes 10 intents with their queries and entities with their lists of synonyms. 
To set up the NLUs, we configure the NLUs to use the list feature for all entities, that is, the NLU will not attempt to extract any entities that are not present in the training set. 
This is thematically in-line with the nature of the Repository task where a chatbot answers questions about software repositories. 
In this context, an entity that does not exist in the repository (e.g., wrong Jira ticket number) is not useful for the chatbot and cannot be used to extract any information for the user. Then, using the NLUs' API, we define the entity types that exist in the Repository corpus, namely 1) CommitHash 2) JiraTicket 3) FileName, and use a fourth built-in entity type (DateTime).

In contrast to the Repository task, there is no original split in the Stack Overflow corpus, as this corpus was not originally intended to be used as training sets for chatbots.  
In fact, we augmented the dataset by manually labeling the intents in all queries, as discussed in Section~\ref{subsec:datasets}. 
Therefore, to evaluate the NLUs' performance on the Stack Overflow task, we use a stratified tenfold cross validation \cite{Han_2011Elsevier}. The cross validation randomly divides the corpus into ten equal folds where nine folds are used for training and one fold for testing.
Since we used the stratified version, the method maintains a consistent distribution of intents across all folds. It is important to note that we use the same process to train and test all the NLUs, hence, the queries remain consistent across the NLUs in each run of the tenfold cross validation.
Moreover, we do not use the list feature to train the NLUs on all entities in the Stack Overflow task. Instead, we train the NLUs to extract the entities using the prediction feature.
This allows us to evaluate the NLUs' ability to extract entities that they have not been trained on before, which better emulates real-life scenarios where practitioners cannot expect to train their chatbots in all possible entities.

We train Dialogflow, LUIS, and Rasa using their APIs, and train IBM Watson using its user interface. 
After the training process, we send each query in the test set through all NLU APIs and analyze their response, which includes the classified intent, the intent's confidence score, and the extracted entities.
This response is then compared against the oracle to evaluate the performance of the NLU. To enable the replicability of our study, we make the scripts used to evaluate the NLUs performance and datasets publicly available~\cite{NLUsCompScript_Link}.

To evaluate the performance of the NLUs in each task, we calculate the standard classification accuracy measures that have been used in similar work (e.g., \cite{braun_SIGDIAL2017}) - precision, recall, and F1-measure. In our study, precision for a class (i.e., intent or entity) is the percentage of the correctly classified queries to the total number of classified queries for that class (i.e., Precision = $\frac{TP}{TP+FP}$). The recall for a class is the percentage of the correctly classified queries to the total number of queries for that class that exist in the oracle (i.e., Recall =  $\frac{TP}{TP+FN}$). Finally, to present the overall performance of each NLU, we use the weighted F1-measure. In particular, we compute the F1-measure (i.e., F1-measure =  \(2 \times \frac{Precision \times Recall}{Precision + Recall}\)) for each class and aggregate all classes F1-measure using weighted average, with the class' support as weights. This approach of calculating the F1-measure has been used in similar work~\cite{Barash_2019FSE,Ilmania_2018IALP}.
While we evaluate all three metrics, we only showcase the weighted F1-measure in the paper, whereas the precision, recall, and F1-measure for each intent and entity are presented in the Appendix.

\section{Case Study Results}
\label{sec:results}

In this section, we present the comparison of the NLUs' performance in terms of intents classification, confidence score, and entity extraction on the Repository and Stack Overflow tasks.

\begin{figure}
	\centering
	\includegraphics[width=\linewidth]{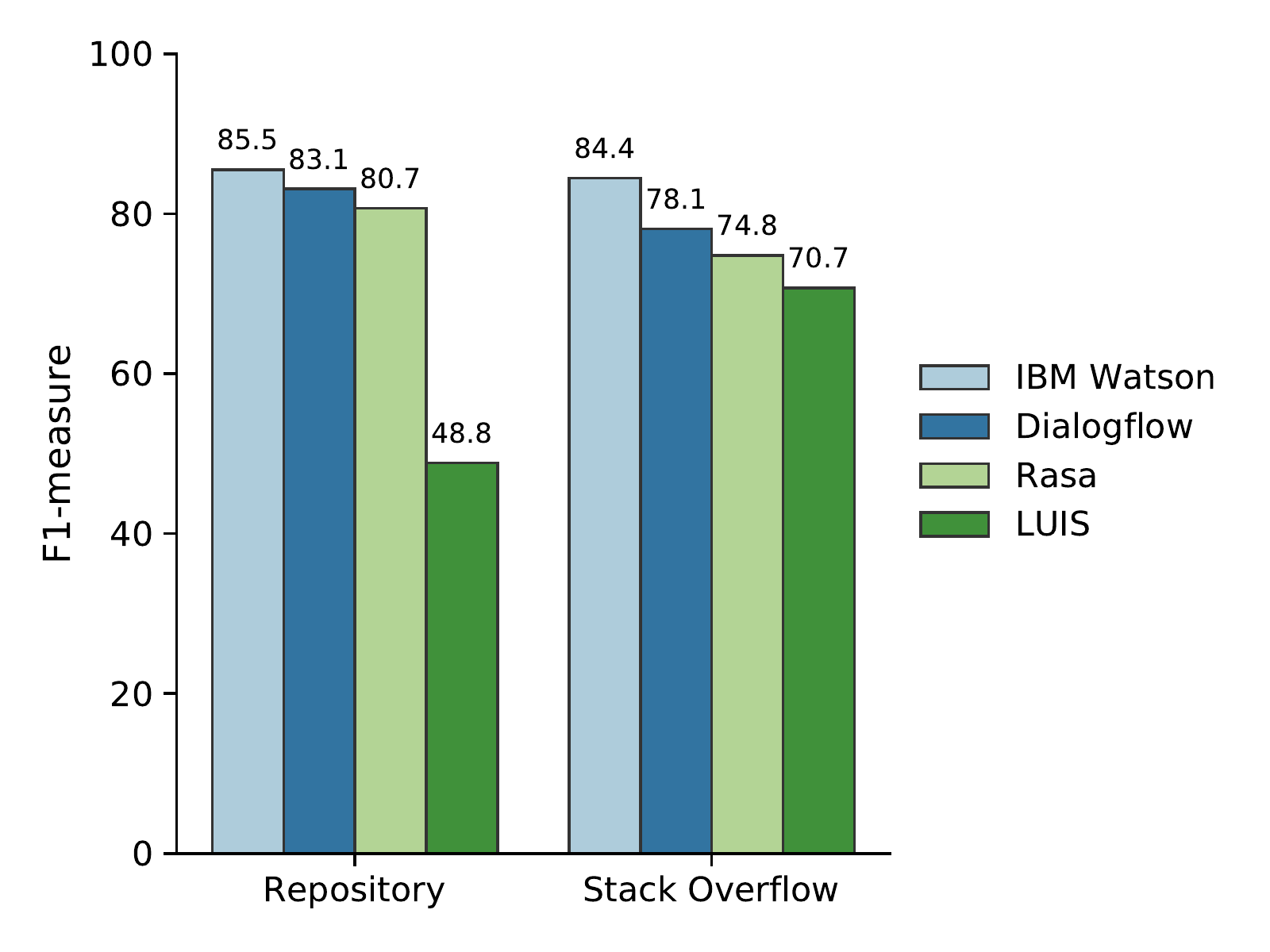}	
	\caption{Intent classification performance as F1-measure of the four NLUs.}
	\label{fig:IntentsClassification}	
\end{figure}

\begin{table*}[tbh]
	\centering
	\caption{Intents' characteristics and classification performance as F1-measure of the four NLUs.}
	\begin{tabular}{l l|c c c|r r r r r}
		
			\toprule
			\multirow{3}{*}{\textbf{Task}} & \multirow{3}{*}{\textbf{Intent}} & \multicolumn{3}{c|}{\textbf{Intent's characteristics}} & \multicolumn{5}{c}{\textbf{F1-measure}} \\
			
			& & \textbf{\#  Training} & \textbf{\% Queries w.} & \textbf{Distinct Entity} & \textbf{IBM} & & & &\\
			
			& 		&  \textbf{Samples} &  \textbf{Exclusive Words} & \textbf{ Types} & \textbf{ Watson} & \textbf{Dialogflow} & \textbf{Rasa} & \textbf{LUIS}&\textbf{Avg.}\\
			
			\midrule
			\multirow{10}{*}{\rotatebox[origin=c]{90}{ \textbf{Repository}}} 
			& \textbf{BuggyFile} & 37 & 92 & - & 96.0 & 96.0 & \textbf{100.0} & 83.3 & 93.8\\
			& \textbf{FixCommit} & 31 & 10 & CommitHash & \textbf{100.0} & 91.7 & \textbf{100.0} & 68.8 & 90.1\\
			& \textbf{BuggyCommit} & 52 & 7 & JiraTicket & \textbf{94.7} & \textbf{94.7} & \textbf{94.7} & 64.3 & 87.1\\			
			& \textbf{OverloadedDev} & 15 & 88 & - & \textbf{94.1} & 80.0 & 88.9 & 58.1 & 80.3\\			
			& \textbf{BuggyCommitByDate} & 66 & 39 & - & 72.7 & 69.6 & 80.0 & \textbf{91.7} & 78.5\\			
			& \textbf{CountCommitsByDate} & 11 & 97 & - & 88.4 & \textbf{91.3} & 80.0 & 53.3 & 78.3\\  
			& \textbf{BuggyFixCommit} & 32 & 74 & - & \textbf{100.0} & 85.7 & 63.2 & 48.0 & 74.2\\ 							
			& \textbf{ExperiencedDevFixBugs} & 15 & 72 & - & \textbf{92.3} & 88.9 & \textbf{92.3} & 13.3 & 71.7\\			
			& \textbf{CommitsByDate} & 8 & 70 & - & 70.0 & 50.0 & \textbf{76.2} & 0.0 & 49.1\\
			& \textbf{FileCommits} & 10 & 77 & - & 55.6 & \textbf{80.0} & 28.6 & 11.1 & 43.8\\
			\midrule
			\multirow{5}{*}{\rotatebox[origin=c]{90}{{\textbf{\shortstack[c]{Stack\\ Overflow}}}}}
			& \textbf{LookingForCodeSample} & 132 & 100 & Platform & \textbf{90.9} & 85.9 & 84.5 & 84.0 & 86.3\\			
			& \textbf{LookingForBestPractice} &12 & 92 & - & 81.3 & 81.3 & 78.0 & \textbf{83.3} & 81.0\\ 			
			& \textbf{UsingMethodImproperly} & 51 & 100 & - & \textbf{79.1} & 66.5 & 64.4 & 57.5 & 66.9\\
			& \textbf{FacingError} &10 & 100 & - & \textbf{80.0} & 60.0 & 33.3 & 10.0 & 45.8\\
			& \textbf{PassingData} & 10 & 70 & - &  36.7 & \textbf{50.0} &  36.7 & 0.0 & 30.9\\
			\bottomrule
		\end{tabular}
		\label{table:IntentsClassificationPerIntent}	
\end{table*}

\subsection{Intents Classification}
\label{subsec:intents_classification_results}

To evaluate the NLUs in intents classification, we train and test each of the NLUs using the corpus from each of the two SE tasks. 
When testing the NLUs, we only consider the top scoring intent as the classified intent for two reasons.
First, to emulate real life use-cases where chatbots use the intent with the highest corresponding score, as it is the intent with the highest probability of being correct~\cite{RasaConfidence_link, LUISConfidence_link, DialogflowConfidence_link, WatsonConfidence_link}.
Second, to ensure that the evaluation of all NLUs is consistent, as Dialogflow only returns one intent with the corresponding confidence score in its response for a single query.
\\

\noindent\textbf{Results.} Figure~\ref{fig:IntentsClassification} shows the F1-measure for intent classification. The Figure presents the performance for each SE task, per NLU. The ranking is consistent across both tasks, showing that IBM Watson outperforms other NLUs, achieving an F1-measure of 85.5\% in the Repository task and 84.4\% in the Stack Overflow task. We also observe that for both SE tasks, LUIS comes last in intents classification.

To obtain a more detailed view of the NLUs' performance, we present the characteristics of all intents (columns 3-5) and the NLUs' F1-measure values (columns 6-9) in Table~\ref{table:IntentsClassificationPerIntent}. We highlight in bold the best performing NLU for each intent. Our results show that IBM Watson is the best performing NLU for 8 intents, followed by Rasa, which performs best for 5 intents.

Moreover, we observe that some intents are more difficult to classify than others. For example, the \CommitsByDate, \FileCommits, \FacingError, and \PassingData intents are difficult to classify, with at least 2 of the four NLUs achieving an F1-measure $<$ 60\%.

Given the performance of the NLUs on the different intents, we inspect the queries of each intent and find three main factors that contribute to the NLUs' varying performance. First, NLUs tend to generally perform well for intents that have a higher number of training samples (e.g., the \LookingForCodeSample intent). Second, NLUs better classify intents that contain queries with exclusive words, that is, words that do not appear in any other intents (see the column \% Queries w. Exclusive words in Table~\ref{table:IntentsClassificationPerIntent}).
For example, \FacingError is the intent with the highest average score among all NLUs. This intent contains exclusive words, such as `fail' and `crash', in all its queries.  
Conversely, the NLUs misclassify intents that more frequently share words with other intents (i.e., those intents have less exclusive words). For example, \PassingData intent has similar \# Training Samples as \FacingError intent but less \% Queries w. Exclusive words. 
Third, intents that have distinct entity types in their queries (e.g., \Platform entity type occurs only with \LookingForCodeSample intent queries) are better classified by certain NLUs since such NLUs use the extracted entity types as input for the intent classification~\cite{EntityIntentExtractionDF_link,EntityIntentExtractionIBMWatson_link}.
This is clearly shown by the high F1-measure yielded by IBM Watson and Rasa in the classification of the intents \FixCommit and \BuggyCommit, which have specific entity types (i.e., \CommitHash and \JiraTicket). %\emad{add NLU names in the last sentence to be specific.}\ahmad{Done.}

\conclusion{
	NLUs rank similarly in both tasks in intents classification, with IBM Watson outperforming all other NLUs, followed by Dialogflow, Rasa, and LUIS.
	Aside from the training sample size, intents that contain exclusive words and distinct entity types are easier to identify by all NLUs.
}

\subsection{NLUs Confidence scores}
\label{subsec:nlusConfidenceScoresResults}

\begin{figure*}
	\centering
	\includegraphics[width=\linewidth]{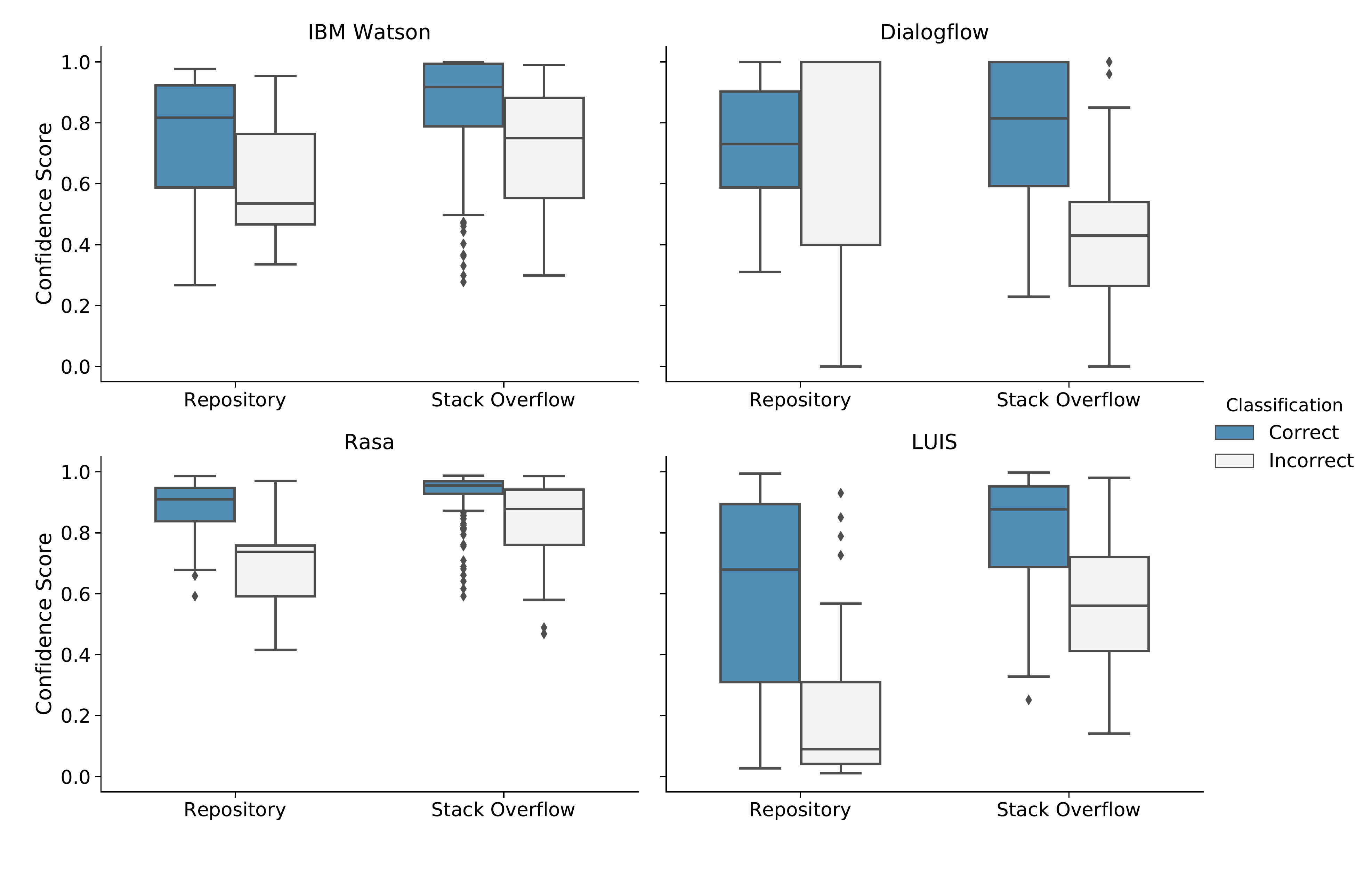}
	\caption{Confidence score distribution for all NLUs and tasks.}
	\label{fig:ConfidenceScoreResults}
	
\end{figure*}

As discussed earlier in Section~\ref{sec:background}, every intent classification performed by the NLU has an associated confidence score. The confidence score is used to determine how much trust one can put into the intent classification of the NLU. Typically, NLUs should provide high confidence scores for intents that are correctly classified. For example, if a query is asking ``What is the number of commits between 1-July-2020 to 1-August-2020?'' and the NLU provides a high confidence score (e.g., 0.98) when attributing the query to the \CountCommitsByDates intent, then the users of the NLU can trust these confidence scores. Also, the contrary is true, if an NLU provides high confidence scores to wrongly classified intents, then one would lose trust in the NLUs' produced confidence scores.

To compare the performance of the NLUs in terms of confidence scores, we queried each NLU with questions from the two tasks - Repository and Stack Overflow. For each query, we considered the highest confidence score to map to a specific intent. For example, for the query above (``What is the number of commits between 1-July-2020 to 1-August-2020?''), an NLU may return a confidence score of 0.98 for the intent \CountCommitsByDates, and a confidence score of 0.80 for the intent \BuggyCommitsByDate. If the correct intent is \CountCommitsByDates, then we would consider this as a correctly classified instance and record the confidence score. In cases where the top confidence score does not indicate the correct intent, we record this an incorrect classification and record the intent confidence score. 

We present the distributions of the confidence scores and compare these distributions for correctly and incorrectly classified intents. Ideally, NLUs should have high confidence scores for correctly classified intents (and vice versa). In addition, having clearly disjoint distributions of the confidence scores between the correctly and incorrectly extracted intents indicates that practitioners can rely on the confidence score of the NLU.

\noindent\textbf{Results.} Figure~\ref{fig:ConfidenceScoreResults} shows the distributions of confidence scores returned by IBM Watson, Dialogflow, Rasa, and LUIS in the Repository and Stack Overflow tasks. From Figure~\ref{fig:ConfidenceScoreResults}, we observe that all NLUs return higher median confidence scores for the correctly classified intents compared to the incorrectly classified intents, for both tasks. The sole exception is Dialogflow, which has a higher median confidence score for incorrectly classified intents for the Repository task. 

Among the evaluated NLUs, Rasa stands out as being the NLU with the highest corresponding confidence scores medians (higher than 0.91) in both the Repository and Stack Overflow tasks, for the correctly classified intents. Furthermore, Rasa has the most compact distribution of confidence scores among other NLUs and the least overlapping confidence scores between the correctly classified and misclassified intents. This means that when Rasa returns a high confidence score, it is highly likely to be correct.

To ensure that the difference in the confidence scores between the correctly and incorrectly classified intents across NLUs is statistically significant, we perform the non-parametric unpaired Mann-Whitney U test on each NLU results. We find that the differences are statistically significant (i.e., p-value $< 0.05$) in all cases and for both, the Repository and Stack Overflow tasks, except for the results of Dialogflow in the Repository task. Generally, our results show that developers can trust the confidence score yielded by NLUs to assess if the NLUs have correctly classified the intent, or the chatbot needs to trigger the fallback action, an action that is used when an NLU cannot determine a correct intent.

As mentioned earlier, the only outlier in our evaluation is the experiment from Dialogflow on the Repository task.
Dialogflow returns a higher confidence score median for the misclassified intents (1.0) compared to the correctly classified intents (0.73), in the Repository task (see Figure~\ref{fig:ConfidenceScoreResults}). We searched the online documentation and forums to see if others have faced similar situations. We found multiple posts where developers raise issues with Dialogflow's high confidence scores for incorrectly classified intents~\cite{chatbotD93:online,DialogFlowIssueConfidenceScore_Link}. This indicates that our results are not an outlier and there might be an issue with Dialogflow that needs to be addressed. The developers reported that they rely on workarounds to overcome such issues, such as combining the confidence score with other measures (e.g., regular expression) \cite{DialogFlowIssueConfidenceScore_Link}.

\conclusion{Overall, NLUs yield higher confidence scores for correctly classified intents. IBM Watson, Rasa, and LUIS provide higher median confidence scores, ranging between 0.68 - 0.96, for correctly classified intents.}

\subsection{Entity Extraction}
\label{subsec:entityExtractionResult}
To correctly answers users' queries, chatbots need to also correctly extract the relevant entities. We consider the extracted entity to be correct only if its type and value exactly match the type and value of expected entity for that query in the oracle. The reason behind our criteria is that extracting entities that are only partially correct (i.e., have only correct values or correct types), causes the chatbot to incorrectly reply to the user's query. Since there can exist a varying number and types of entities in each query, we need a mechanism to ensure that our evaluation accounts for such variance. To mitigate this issue, we calculate the precision, recall, and weighted F1-measure based on the number of entities in each entity type, for the entity extraction results.

\begin{figure}[tbh]
	\centering
	\includegraphics[width=\linewidth]{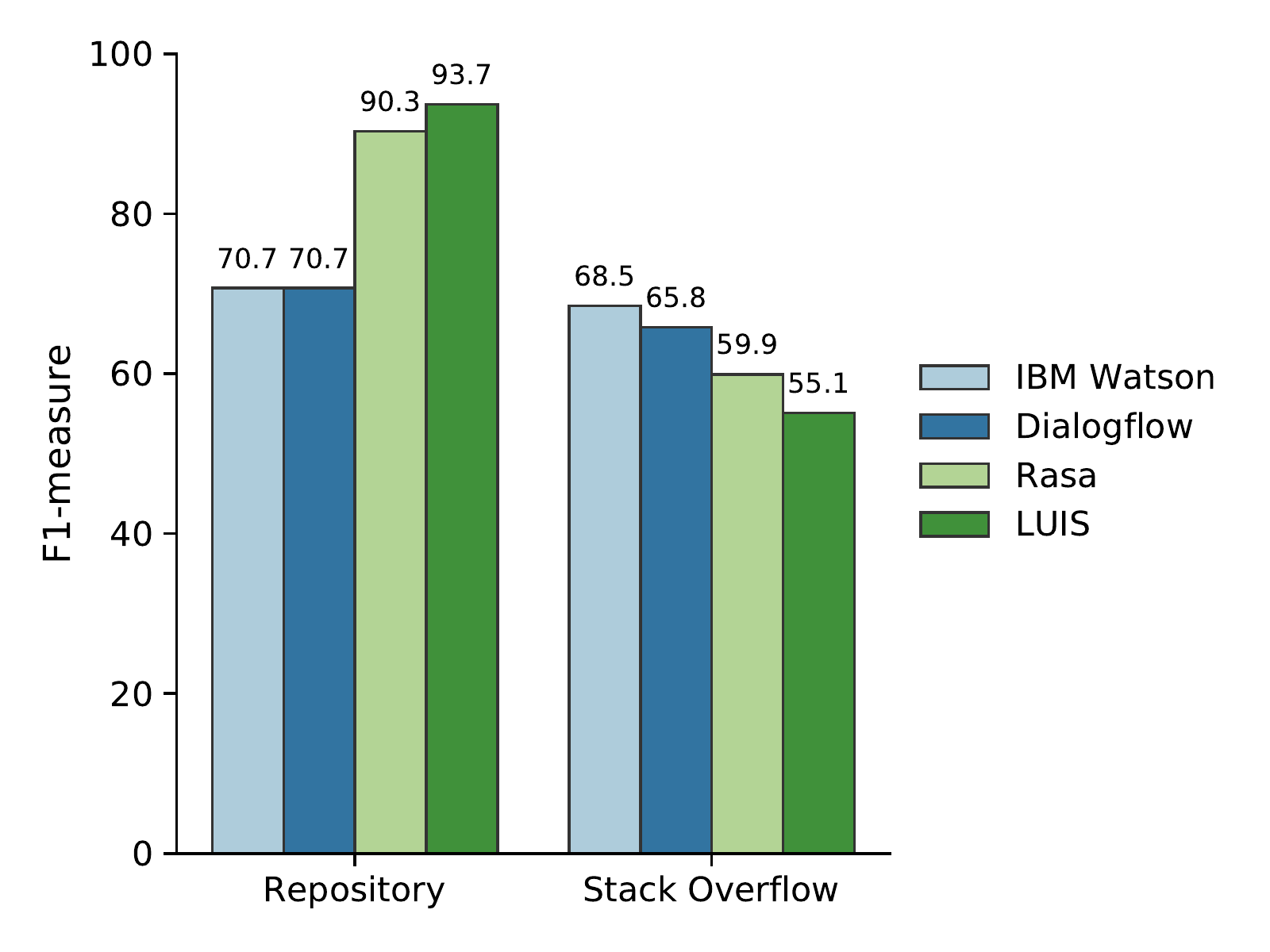}
	\caption{Entity extraction performance as avg. F1-measure of the four NLUs.}
	\label{fig:EntityExtractionPerformance}
\end{figure}

\noindent\textbf{Results.} 
Figure~\ref{fig:EntityExtractionPerformance} presents the entity extraction results for the two tasks. To better interpret the results, we reiterate that extracting entities in the Repository and Stack Overflow tasks require different NLU features. 
For the Repository task, we configure the NLUs to extract entities using the list feature (i.e., list of synonyms) and entities are extracted as an exact match.
In the Stack Overflow task, however, we configure the NLUs to use the prediction feature as discussed in Section~\ref{subsec:datasets}, that is, requesting the NLUs to predict the correct entity. 
Given their differences, we discuss the results of each task separately, first describing the results for the Repository task and then the Stack Overflow task.% \emad{refer to Section 3.2}\ahmad{Done.}

\noindent\textbf{Repository.} According to Figure~\ref{fig:EntityExtractionPerformance}, LUIS is the best performing NLU with an average F1-measure of 93.7\%, followed by Rasa (90.3\%). Both IBM Watson and Dialogflow perform similarly (70.7\%) when extracting entities in the Repository task.

To better understand the factors influencing the performance, we examine the F1-measure of the NLUs in light of the different entity types. Table~\ref{table:EntityExtractionPerEntity} shows the NLU's performance per entity type.
We highlight, in bold, the best performing NLU for each entity type.
Our results confirm that LUIS is the best performing NLU when extracting the \CommitHash, \JiraTicket, and \FileName entity types from the Repository task. IBM Watson also performs well in the \CommitHash and \JiraTicket entity types, but it falls short when extracting  \FileName entities (F1-measure of 15.9\%).
This is due to IBM Watson's inability to differentiate between the normal words (e.g., `To' and `A') and entities of type \FileName (e.g., `To.java' and `A.java').
In other words, it extracts entities based on their exact match with the training set without considering the entities' context when using the list feature \cite{IBMWatsonEntitiesSimilarity_link}.
For \DateTime entities, Rasa performs best in extracting all the dates correctly (F1-measure of 100\%) from the given queries. Rasa uses a probabilistic context-free grammar through their pipeline (Duckling) to extract the dates from the posed queries \cite{Duckling_link}. 
For other NLUs, we notice two reasons behind the incorrect extraction of the \DateTime entities: 1) the misspelling of the date from the users (e.g., ``what are the commits I submit on 27/5/2018 $\sim$ 31/5/2018'') 2) vague date formats in queries (e.g., ``Tell me about the commits from 05-27 to 05-31 in 2018'').

\noindent\textbf{Stack Overflow.} 
We observe in Figure~\ref{fig:EntityExtractionPerformance} that IBM Watson yields the best results (F1-measure of 68.5\%), followed by Dialogflow (65.8\%), Rasa (59.9\%), and LUIS (55.1\%).
Note that the performance of the NLUs on the Stack Overflow task is expectedly lower than the performance obtained in the Repository task, given that NLUs have to predict entities in a query.

Similar to the case of the Repository task, we also examine the performance of the NLUs when extracting the different entity types in the Stack Overflow task. Table~\ref{table:EntityExtractionPerEntity} shows that the performance of the NLUs varies from one entity type to the other, and that no NLU outperforms the rest in extracting the majority of entity types. Table~\ref{table:EntityExtractionPerEntity} shows that, in the Stack Overflow task, both IBM Watson and Dialogflow outperform other NLUs in extracting two different entity types. Furthermore, we observe that some entity types are more difficult to extract than others. In particular, entities of type \Framework, \Standards, and \API are difficult to extract (i.e., the four NLUs achieve an F1-measure $<$ 60\%).
Upon closer investigation of the results, we find that more than 60\% of the most difficult entities appear only once in the task, that is, they are unique entities. %in the dataset. 
For example, the entity `DOM' of type \Standards in the query ``How to tell if an element is before another element in DOM'' is a unique entity as it occurs only once in the Stack Overflow task. This entity was not extracted by any of the evaluated NLUs. Consequently, NLUs tend to perform well when extracting entities which appear frequently in the training set (i.e., not unique).

\conclusion{
Overall, NLUs perform differently in the two evaluated tasks, with LUIS and Rasa outperforming others when using the list feature (Repository task) for entity extraction, while IBM Watson and Dialogflow perform better when the entities need to be predicted (Stack Overflow task).
}

\subsection{Concluding Remarks}
\label{subsec:ConcludigRemarks}
In the previous three sections, we compare the NLUs' performance in terms of three aspects (intents classification, confidence scores, and entity extraction). 
We find that the performance of each NLU can vary from one aspect to another, and between the two tasks. In other words, there is no NLU that outperforms all others in every aspect. Hence, in this section, we set up to rank the NLUs on their overall performance (considering the results from all aspects studied in the previous sections) in order to find the best NLU for chatbots in the SE domain. Having an overall measure is challenging since we have different tasks (Repository and Stack Overflow) and are using different entity extraction features (i.e., list vs. prediction features). Therefore, we use an approach that has been used in prior work~\cite{Shihab_2011SPC} to rank techniques evaluated against different datasets. In particular, we use F1-measures of each task for intents classification (Section~\ref{subsec:intents_classification_results}) and entity extraction (Section~\ref{subsec:entityExtractionResult}) aspects. For the confidence score aspect, we rank the NLUs using the median confidence scores for the correctly classified intents of each task as shown in Section~\ref{subsec:nlusConfidenceScoresResults}. To rank the NLUs, we compute the NLUs' average rank using their ranks in all aspects and tasks. The NLU with the lowest average rank is the best performing NLU.

\begin{table}[]
	\centering
	\caption{Entity extraction performance as F1-measure per entity of the four NLUs.}
	\label{table:EntityExtractionPerEntity}
	\addtolength{\tabcolsep}{-2.5pt}    	
	\begin{tabular}{l l rrrrr}
		\toprule
		\multicolumn{1}{c}{\multirow{2}{*}{\textbf{Task}}} & \multicolumn{1}{c}{\multirow{2}{*}{\textbf{Entity Type}}} & \multicolumn{5}{c}{\textbf{F1- measure}} \\
		& \multicolumn{1}{c}{} & \textbf{IBM Watson} & \textbf{Dialogflow} & \textbf{Rasa} & \textbf{LUIS} & \textbf{Avg.} \\
		\midrule
		\multirow{4}{*}{\rotatebox[origin=c]{90}{ \textbf{Repository}}} 
		& \textbf{CommitHash} & \textbf{100.0} & \textbf{100.0} & 88.9 & \textbf{100.0} & 97.2\\ [0.025cm]	     
		& \textbf{JiraTicket} & \textbf{100.0} & 91.7 & 90.0 & \textbf{100.0} & 95.4\\ [0.025cm]
		& \textbf{DateTime} & 86.2 & 56.3 & \textbf{100.0} & 98.1 & 85.2\\ [0.025cm]
		& \textbf{FileName} & 15.9 & 79.2 & 71.4 & \textbf{80.0} & 61.6\\ [0.025cm]
		\midrule
		\multirow{5}{*}{\rotatebox[origin=c]{90}{{\textbf{\shortstack[c]{Stack\\ Overflow}}}}}
		& \textbf{ProgLanguage} & 92.0 & \textbf{93.7} & 91.4 & 86.8 & 91.0\\
		& \textbf{Platform} & 67.4 & 55.9 & \textbf{75.1} & 43.6 & 60.5\\ 
		& \textbf{Framework} & \textbf{65.4} & 56.0 & 56.1 & 54.4 & 58.0\\
		& \textbf{Standards} & 54.1 & \textbf{56.9} & 14.9 & 17.5 & 35.9\\
		& \textbf{API} & \textbf{43.3} & 42.8 & 29.9 & 23.5 & 34.9\\
		\bottomrule
	\end{tabular}

\end{table}

\newcommand{\ra}[1]{\renewcommand{\arraystretch}{#1}}
\newcommand{\grayrow}{\rowcolor{gray!10}}
\ra{1.1}
\begin{table*}[bth]
	\centering
	\caption{NLUs' overall performance ranking. }
	\label{tab:NLURanks}
	\makebox[\linewidth]{
		\begin{tabular}{l|>{\centering}p{1.5cm}>{\centering}p{1.3cm}>{\centering}p{1.3cm}|>{\centering}p{1.5cm}>{\centering}p{1.3cm}>{\centering}p{1.3cm}|p{0.65cm}}
			\toprule
			\multirow{2}{*}{} & \multicolumn{3}{c|}{\textbf{Ranking in Repository Task}} & \multicolumn{3}{c|}{\textbf{Ranking in Stack Overflow Task}} & \multirow{2}{*}{}  \\
			\rule{0pt}{0.3cm} 				\textbf{NLU}   & \textbf{Intents\newline Classification}  & \textbf{Confidence\newline Score} & \textbf{Entity\newline Extraction} 						& 	\textbf{Intents\newline Classification} & \textbf{Confidence\newline Score} & \textbf{Entity\newline Extraction} & \textbf{Avg.\newline Rank} \\
			\midrule 
			
			IBM Watson &   1 & 2 &  \hspace{0.1cm}3* & 1 & 2 & 1 &  1.7 \\
			Rasa & 3 & 1 & 2 & 3 & 1 & 3 &  2.2 \\
			Dialogflow 	& 	2 & 3 &  \hspace{0.1cm}3* & 2 & 4& 2 & 2.7 \\
			LUIS & 4 & 4 & 1 & 4 & 3 & 4 &  3.3 \\
			
			\bottomrule
	\end{tabular}}
	\begin{tablenotes}
		\small
		\addtolength{\itemindent}{2.2cm}
		\item 	* Same rank
	\end{tablenotes}
	
\end{table*}
Table~\ref{tab:NLURanks} presents the NLUs' ranks in the different aspects on both tasks and their overall ranks. We find that IBM Watson is the best performing NLU when considering all aspects in both tasks, followed by Rasa, Dialogflow, and LUIS. That said, chatbot practitioners need to consider the most important aspect (i.e., intents classification or entity extraction) and tasks performed by their chatbot when selecting the NLUs they want to use.

\begin{figure*}[tbh]
	\centering		
	\begin{subfigure}{0.5\textwidth}	
		\centering	
		\includegraphics[width=.9\linewidth]{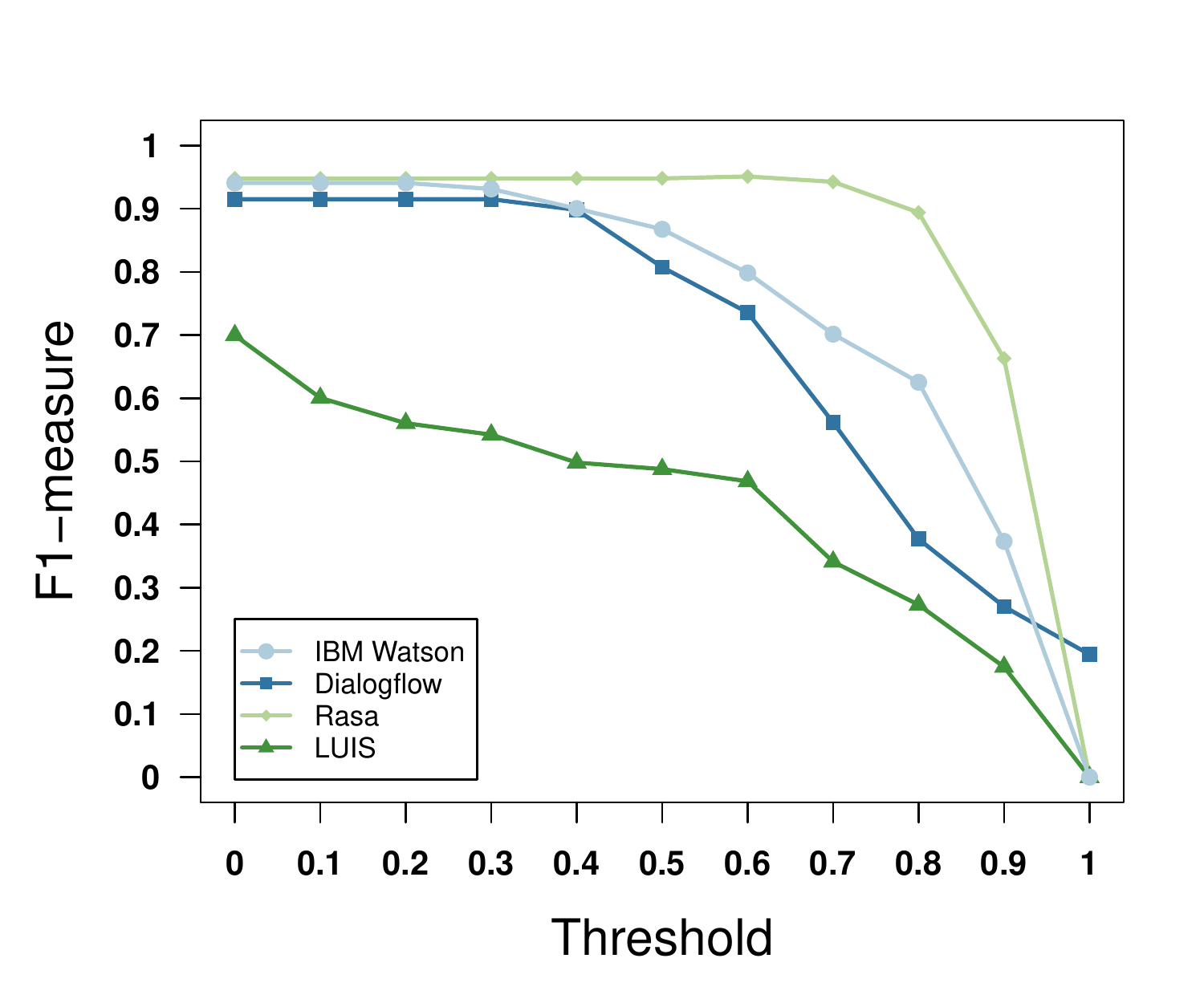}	
		\caption{Ranking in Repository Task}	
	\end{subfigure}\hfill	
	\begin{subfigure}{0.5\textwidth}		
		\centering	
		\includegraphics[width=.9\linewidth]{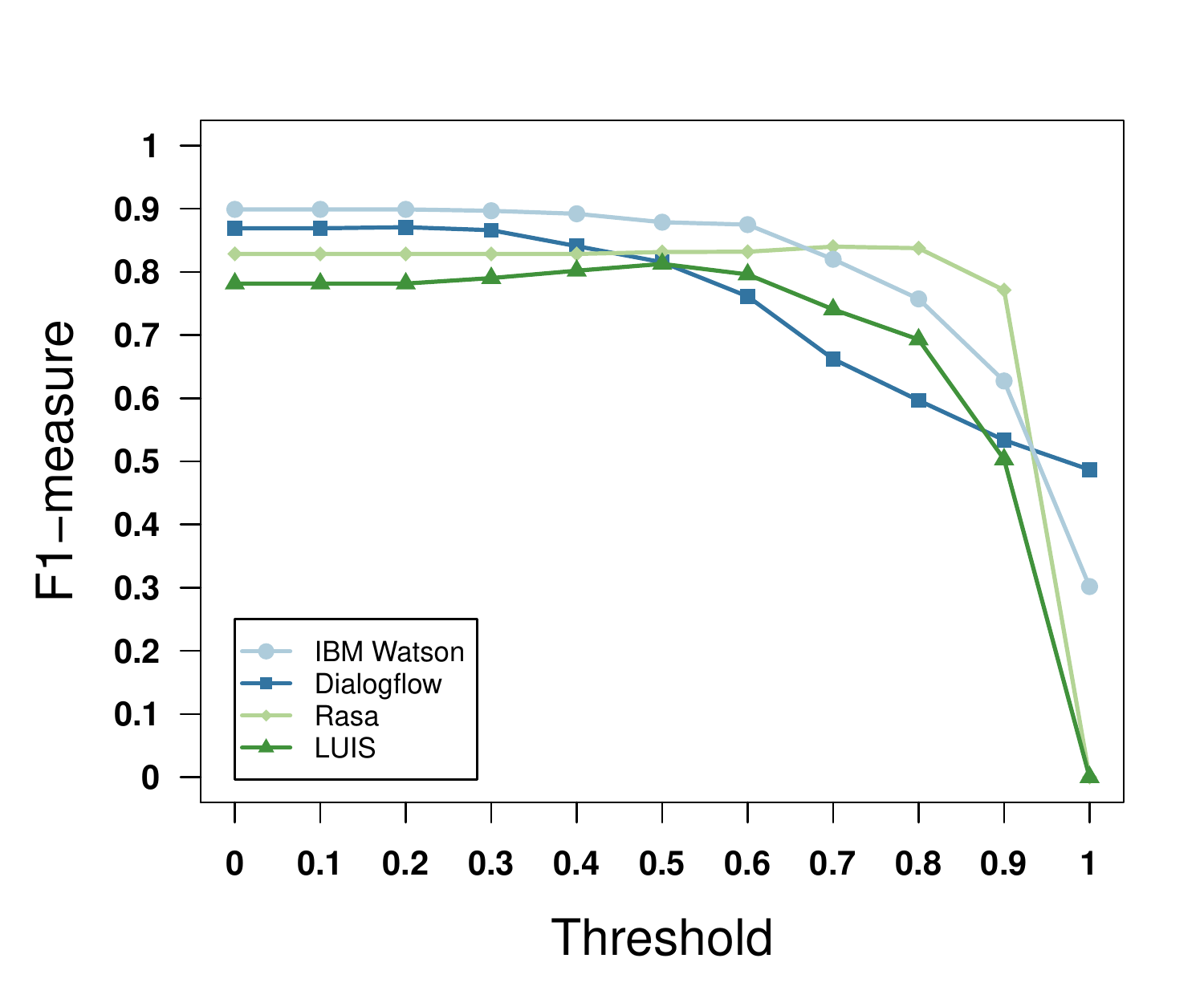}	
		\caption{Stack Overflow Task}	
	\end{subfigure}\hfill	
	\caption{Analysis of theshold sensitivity in terms of F1-measure of the NLUs in the Repository and Stack Overflow tasks.}	
	\label{fig:ConfidenceScoresResults}	
\end{figure*}

\section{Discussion}
\label{sec:discussion}
In this section, we dive into the evaluation results to gain more insights on the NLUs' confidence score sensibility as well as quantifying their abilities to extract unique entities.
Finally, we provide a set of actionable recommendations to chatbot developers and researchers to achieve better intents classification and entity extraction results.

\subsection{Examining the Impact of the Confidence Score Threshold on NLU Performance}
\label{sec:discussion_cs_threshold}

One important impacting factor of the results from these NLUs is the confidence score threshold, i.e., at what confidence score value would an NLU be confident in classifying the returned intent as correct. In some cases, NLUs return high confidence scores for the incorrectly classified intents as shown in Section~\ref{subsec:nlusConfidenceScoresResults}.
Hence, defining the appropriate threshold to accept the classified intent or trigger the fallback action is still a challenge for chatbot developers. In essence, developers have to arbitrarily determine which confidence score value they will use to determine a correct classification.
Given that the confidence score threshold impacts the NLUs' performance, an important question is, how does the confidence score threshold impact the performance?

To maintain consistency with the results presented in Section~\ref{sec:results}, we use the same experimental settings. We vary the confidence score used by each NLU to determine the correct intent and plot its F1-measure performance. 

Figure~\ref{fig:ConfidenceScoresResults} shows the NLUs' F1-measures at varying confidence score thresholds. We perform this analysis for both tasks. For the Repository task, our results show that Rasa is the most robust, achieving consistently high performance for threshold values between 0-0.7. On the other hand, IBM Watson and Dialogflow start to see a reduction in performance for threshold values higher than 0.3. LUIS seems to generally follow a downward trend with higher confidence score thresholds. 
In the case of the Stack Overflow task, we again observe stable performance by Rasa for confidence score thresholds between 0-0.8, while IBM Watson, Dialogflow, and LUIS decrease in performance as the confidence score threshold increases.

From the results obtained, we observe that 1) the NLU's performance varies based on the task at hand and 2) that using lower confidence score threshold values tends to achieve better performance. This is because the selected threshold changes the NLU's performance only when it surpasses some of the confidence scores returned by that NLU. For example, Rasa has a high confidence score median (0.91), and hence, it will only be affected when the threshold increases significantly (threshold $>$ 0.7). That said, one needs to be careful since using a low confidence score may also lead to the chatbot providing the wrong answer to the question being asked. Overall, we recommend that developers should investigate and carefully choose a proper confidence score threshold since such thresholds can have a significant impact on the chatbot's performance.

\subsection{Unique Entities}
The performance of the NLUs in extracting entities from the Stack Overflow task was affected by unique entities, as shown in Section~\ref{subsec:entityExtractionResult}. As the name suggests, unique entities appear just once in the dataset for the Stack Overflow task; thus, the NLUs have to predict their occurrences without prior training. It is important to note that there are no unique entities when evaluating the NLUs using the list feature because the NLUs have been trained on all entities that exist in the Repository task.
To better understand the NLUs' ability to extract unique entities, we investigate the results from the Stack Overflow task, examining the NLU performance on queries containing only unique entities.
Hence, queries containing any non-unique entity were excluded from this investigation.
We find 58 queries that fit our criteria, and they include a total of 75 unique entities that are distributed, as shown in Table~\ref{table:Unique_Entity}. 
Similarly to the evaluation conducted in Section~\ref{subsec:entityExtractionResult}, we calculate the precision, recall, and weighted F1-measure of the NLUs when extracting unique entities.

Table~\ref{NLU_Extract_Unique_Entities} presents the NLUs' performance in extracting unique entities. We find that IBM Watson outperforms other NLUs with an F1-measure of 31.6\% in extracting unique entities. We examine the results of extracting unique entities and find two factors that impact the NLUs' performance. First, the NLUs depend on the entities syntax similarity to recognize the entities in the queries of the testing set. For example, two NLUs (IBM Watson and LUIS) incorrectly extracted `Receiving 0.0' as a \Framework entity from the query ``Receiving 0.0 when multiplying two doubles in an array'' because `Receiving 0.0' is syntactically similar to some other \Framework entities such as `Spring 4.0.2' and `CodeIgniter 2.0.3'. 
Second, we find that the NLUs extract one of the words in the multi-worded entity (e.g., `SAP crystal reports' is a single entity of type \Framework) as a separate entity on its own, given that the NLUs are trained on such separate entities. For example, the multi-worded entity `Python NameError' with an \API type is extracted by all NLUs as the entity `Python' of type \ProgLanguage. In fact, the multi-worded entities in the Stack Overflow task are all unique entities, except for two entities (i.e., `Internet Explorer' and `WebSphere 8.5.5').
Hence, these results motivate the need for tools/techniques from the research community to extract the unique entities. Also, we encourage the chatbot developers to give special attention to unique entities by training the NLU on more examples by including those entities.

\begin{table}[]
	\centering
	\caption{Distribution of unique entities by entity type in the Stack Overflow task.}
	\label{table:Unique_Entity}
	\begin{tabular}{lrr}
		\toprule
		\textbf{Entity Type} & \textbf{Unique Entities (\%)} \\ 
				
		\midrule
		\textbf{ProgLanguage} & 1 (1.3) \\
		\textbf{Framework} & 36 (48) \\
		\textbf{Standard} & 5 (6.7) \\
		\textbf{API} & 32 (42.7) \\
		\textbf{Platform} & 1 (1.3) \\ 
		\bottomrule
	\end{tabular}
\end{table}

\begin{table}[]
	\centering
	\caption{Precision, Recall, and F1-measure of extracting unique entities in the Stack Overflow task.}
	\label{NLU_Extract_Unique_Entities}
	\begin{tabular}{lrrr}
		\toprule
		\textbf{NLU} & \textbf{Precision} & \textbf{Recall} & \textbf{F1-measure} \\ 
		\midrule
		\textbf{IBM Watson} & 51.4\% & 28\% & 31.6\% \\
		\textbf{Dialogflow} & 0\% & 0\% & 0\% \\
		\textbf{Rasa} & 12\% &1.3\% & 2.1\% \\
		\textbf{LUIS} & 25.3\% & 5.3\% & 7.7\%\\ 
		\bottomrule
	\end{tabular}
\end{table}

\subsection{Recommendations}

\begin{table*}[tbh]
	\caption{Recommendations for fine-tuning the NLUs when developing Chatbots.}
	\label{tab:recommendations}
	
	\begin{tabularx}{\linewidth}{llX}
		\toprule
		\textbf{Problem} & \textbf{Recommendation} \\
		\midrule
		\multirow{3}{*}{Low accuracy on intents classification} 
		& R1. Train NLUs with multiple queries per intent	\\
		& R2. Merge intents with similar words and disambiguate with a follow-up action				\\
		& R3. Combine extra factors (e.g. regex, entity type) to aid to the confidence score \\
		
		\midrule
		
		\multirow{2}{*}{Low accuracy on entity extraction}
		& R4. Use different entity features (list feature vs prediction feature) according to the entity type \\
		& R5. When using entity prediction, focus on including entities in different query positions \\

		\bottomrule
	\end{tabularx}
	
\end{table*}

Based on our findings and experience in conducting this study, we provide a set of actionable recommendations to help chatbot practitioners to improve the performance of the used NLU. 
Table~\ref{tab:recommendations} summarizes our recommendations to improve the NLUs' performance in intents classifications and entity extraction.
While our results are based on SE tasks, some of the guidelines can be used to improve the NLU performance regardless of the domain. 
We discuss the recommendations in details in the following.

\noindent\textbf{R1. Train NLUs with multiple queries per intent.}
Our results show that the NLUs perform better when classifying queries with intents that have more training examples. 
While more data is typically better in any machine-learning task, the focus on crafting a good training set needs to be put in the diversity of queries per intent. 
In fact, some NLUs recommend that each intent has 10 training examples or more~\cite{Frequent42:online} that represent different ways of querying that intent  (e.g., having different sentence lengths~\cite{multiple17:online}). 
NLUs like Dialogflow and LUIS provide an interactive GUI~\cite{HistoryD17:online} to allow the chatbot developers to edit and add training examples using the questions users pose to the chatbot. 
Chatbot developers should leverage this feature, especially at the early stages of the chatbot development, as a part of the debugging process of the NLU and as a way of fine-tuning the initial training set of the chatbot. We plan (and encourage others) to explore semi-supervised learning~\cite{Zhu_2009} and weak supervision~\cite{Zhou_2017NSR} approaches to automate the NLU's retraining process.
\\

\noindent\textbf{R2. Merge intents with similar words and disambiguate with a follow-up action.}
The results show that NLUs better classify intents that contain exclusive words and distinct entity types, not shared with other intents. 
Hence, it is worth exploring the possibility of merging similar intents that share many common words and entity types into a single intent, as the NLUs can misclassify these intents. The merge of intents can be done in two ways: 1) during training, combining queries of similar intents into one intent, or 2) after the intents classification as a chatbot post-processing phase. Option 1 is only recommended when frequently misclassified intents have fewer training examples, as the merging can help boost the initial training dataset. Otherwise, option 2 is a more generally applicable solution as it does not introduce any noise by modifying the training dataset. Once the NLU classifies the merged intent, chatbot developers can employ some strategies to disambiguate the merged intent. The disambiguation can be done using entity extraction, regular expressions, or even relying on follow-up chatbot questions to help extract the target intent. For example, if there is a chatbot that does code refactoring, then the intents `refactor class' and `refactor method' have very similar training examples. In this case, the chatbot developers can merge both intents into one intent (e.g., `refactor'). Then, to identify the target intent (class or method), the chatbot could ask the user about the type of refactoring to perform.
\\

\noindent\textbf{R3. Combine multiple factors to aid the confidence score.}
While most of the NLUs return high confidence scores when correctly classifying intents, as we discussed in Section~\ref{subsec:entityExtractionResult}, we recommend the chatbot developers to use other measures to ensure the intent was classified correctly. For example, chatbot developers can combine the confidence scores with regular expressions to check if specific keywords appear in the question. Another example is an intent that has a distinct entity type where developers can check whether an entity of that type exists in the query and use that information to make sure that the intent was correctly classified (e.g., only queries of `\BuggyCommit' intent have `\JiraTicket' entities).
\\

\noindent\textbf{R4. Use different entity features according to the entity type.}
Developers should resort to different entity features (either list feature or prediction feature) together in the same chatbot, depending on the entities of their domain. 
Entities that are enumerable (e.g., months of the year) and have known finite values are better identified with the features containing all known synonyms.
We observe this when evaluating the entity extraction performance in the Repository task, where NLUs extracted entities more accurately.
However, in most cases, an entity could be expressed in various (and unknown) ways and developers need to resort to the prediction feature (e.g., \Framework entities).
\\

\noindent\textbf{R5. When using the entity prediction feature, focus on including entities in different positions within the queries.}
When it comes to extracting entities using the prediction feature, developers need to diversify their queries so that they include entities in different positions (i.e., beginning, middle, and end of the query) in order to improve the NLU's ability to  extract the entities in different contexts \cite{RASA_ner_crf_Features_code,Goodexam67:online}. Also, developers should expose the NLU with different variants of the entity (e.g., `issues' and `bugs')~\cite{Goodexam67:online}. Finally, we believe that there is a need for a benchmark that contains SE terms (i.e., SE thesaurus) and their variations (e.g., `commit' and `change') as this helps the developers of SE chatbots to train the NLUs on SE entities and their synonyms.

\section{Related Work}
\label{sec:relatedwork}
Since the goal of this paper is to examine the different NLUs for SE chatbots, we discuss the related work in two areas: work that studies chatbots in the SE domain and work that focuses on evaluating NLUs. 

\subsection{Using Chatbots to Assist Developers}
Storey and Zagalsky defined bots as tools that perform tedious tasks to increase developer' productivity and save their time~\cite{Storey_2016FSE}. They outlined five roles for bots in the software development process; coding, testing, DevOps, support and documentation. For example, BugBot~\cite{jshjohns85:online}, a code bot that allows developers to create bugs easily through Slack. Moreover, there are bots that help in code refactoring~\cite{Wyrich_2019BotSE}, recommend reviewers for pull-requests~\cite{Kumar_2019BotSE}, and generate bug fixes~\cite{Simon_2018ICSE}.
\noindent
Chatbots are a sub-category of software bots~\cite{Lebeuf_2019BotSE,lebeuf2018taxonomy} where they are getting more attention from the practitioners in the SE domain \cite{ Lebeuf_2018IST, Bradley_2018ICSE, Toxtli_2018CHI,Beschastnikh_2017ASE, Paikari_2019BotSE, Tian_2017ASE, Urli_2018ICSE-SEIP, Abdellatif2019EMSE}. For example, \citet{Toxtli_2018CHI} developed TaskBot using LUIS, which helps developers within teams to manage their tasks by initiating them and tracking reminders. \citet{Abdellatif2019EMSE} leveraged Dialogflow to implement the MSRBot to help answer developers' questions related to their software projects. \citet{Bradley_2018ICSE} developed Divy using Amazon Alexa to assist developers in their daily development tasks. Moreover, chatbots are used to resolve code conflicts~\cite{Paikari_2019BotSE} and answer development questions using Stack Overflow~\cite{Romero_2020BotSE,Xu_2017ASE,Murgia_2016CHI}.

The mentioned studies help motivate our work; to guide developers and researches to select an NLU for their chatbots based on their needs and the context of usage in the SE domain. However, our work differs in that we evaluate different NLUs and we do not propose new chatbots. \newline 

\subsection{Evaluating and Comparing NLUs}
Recently, many researchers compared different NLUs from different perspectives \cite{angara_2018towards, canonico_CC2018, Shawar_2007NAACL}. For example, \citet{canonico_CC2018} assessed the different aspects/features of NLUs, such as usability, supported programming languages, and the ability to integrate with third parties. Also, the authors tested different NLUs using a dataset from the weather domain. The results show that IBM Watson outperforms other NLUs in terms of classifying the correct intents. Also, \citet{angara_2018towards} compared different attributes of NLUs such as security to help users select a suitable NLU based on the attributes of the chatbot. \citet{gregori_2017GIT} evaluated different NLUs using frequently asked questions on a students admission corpus. The results show that LUIS outperforms other NLUs. The work closest to ours is the work by \citet{braun_SIGDIAL2017}, which evaluated different NLUs using public transport connections, Ubuntu platform, and web applications corpora. The results showed that LUIS achieved the best compared to other NLUs across all datasets. The main difference between the work by Braun et al. and ours is that we evaluate the NLUs using different features (i.e., list and prediction features) on SE tasks rather than technical support tasks (e.g., setup printers).

Our work differs and complements the prior work in two ways. First, our work uses data from two common tasks that are highly relevant to the SE domain, as we focus on providing insights to developers of SE chatbots. Second, our performance evaluation covers different NLU features (e.g., list feature) and pinpoint the impact of some characteristics of entities in the performance of NLUs, which are not previously studied by prior work. 
Overall, our work complements the studies evaluating NLUs and contributes to that body of research by providing a thorough evaluation of NLUs in the SE domain.

\section{Threats to validity}
\label{sec:threats}
In this section, we discuss the threats to internal, construct, replicability, verifiability, conclusion, and external validity of our study. 

\subsection{Internal Validity}
Concerns confounding factors that could have influenced our results. The Stack Overflow tasks used to evaluate the NLUs' performance were manually labelled which might introduce human bias. To mitigate this threat, we had multiple annotators to label the tasks and used the discussion and voting to resolve all disagreements between the annotators. Another threat to internal validity is that we use the default configurations for all NLUs in our study, which could impact their performance. For example, we use the default NLU confidence score threshold to present our results in Section~\ref{sec:results}, but mitigate this threat by studying the impact of confidence score threshold on the NLUs' performance in Section~\ref{sec:discussion_cs_threshold}. 
Other parameters could be tuned to enhance the NLUs' performance (spell-correction and training validation~\cite{AgentValidation_link}). 
We purposely decided to evaluate the NLUs under their default configurations, to evaluate the performance of an average user would encounter when deploying the NLUs in a chatbot. Also, the only common configuration across all NLUs is the confidence score threshold. In other words, some NLUs have configurations that could be tuned, which are not available in the other NLUs or have been implicitly defined in the NLU (since their internal implementation is closed source). Hence, modifying these configurations in one NLU might lead to different results and conclusions.

\subsection{Construct Validity}
Considers the relationship between theory and observation, in case the measured variables do not measure the actual factors. To evaluate the NLUs' performance in the Repository task, we use the MSRBot corpus that was created to evaluate the MSRBot. The MSRBot dataset might have some limitations, such as having questions (intents) that might be less popular than others in real settings. However, we argue that the questions that MSRBot supports were derived via a semi-structured process that collected the most common questions asked by software practitioners from previous studies~\cite{Begel:2014:ATQ:2568225.2568233,Sharma:2017:DWA:3100317.3100333,Fritz:2010:UIF:1806799.1806828}. On the other hand, the MSRBot evaluation participants were free to word their questions to the chatbot. Finally, the list of questions used to train the MSRBot was not revealed to the participants.

\subsection{Replicability Validity}
Concerns the possibility of replicating the study~\cite{Epskamp_2019}. In our study, we used three of the most popular NLUs that are closed source (except for Rasa). The NLU internal implementation might change without any prior notice to the users. Therefore, replicating the study might lead to different results. We mitigate this issue by providing the scripts and datasets used in each task and the version used for Rasa (open-source). We encourage the scientific community to replicate the study through our replication package~\cite{NLUsCompScript_Link} after a certain time (e.g., 6 months) to examine if there is a change in the NLUs performance.
We believe that our study presents a snapshot of the NLUs' performance comparison in the SE domain. Moreover, our results serve as a starting point for the chatbot practitioners when selecting the NLUs to use.

\subsection{Verifiability Validity}
Concerns the verifiability of the study results~\cite{brundage2020trustworthy}. In our study, we compared the performance of different NLUs using two common SE tasks (i.e., Repository and Stack Overflow tasks). Using different NLUs or different tasks might lead to different conclusions. To mitigate this threat, we selected NLUs that are used in prior work~\cite{braun_SIGDIAL2017, gregori_2017GIT, koetter_2018ICART} as a first step to benchmark the NLUs in the SE domain. Also, we described our case study setup in Section~\ref{sec:casesetup}, studied each task characteristics and detailed our analysis and results in Section~\ref{sec:results}. Finally, we shared the dataset used to train/test NLUs, NLUs' responses~\cite{NLUsCompScript_Link}, and the used scripts~\cite{NLUsCompScript_Link} with the community to allow for further investigation and accelerate the future research in the field.

\subsection{Conclusion Validity}
Concerns the relationship between the treatment and the outcome. In Section~\ref{subsec:ConcludigRemarks}, we compute the overall NLUs' performance in both tasks by ranking the NLUs on each task based on their F1-measure then compute their average ranks in all tasks. The NLU's performance differs based on its intended usage (i.e., intents classification and entity extraction) and task (i.e., Repository and Stack Overflow). The main goal behind this analysis is to find the NLU that could be the best to use in the initial implementation of the SE chatbots. Moreover, this analysis has been used in similar studies~\cite{Shihab_2011SPC} to obtain an overall performance for the NLUs. Finally, our results show consistency in the NLUs' ranks in all usages (except for entity extraction) for both tasks.

\subsection{External Validity}
Concerns the generalization of our findings. While we use four of the most commonly used NLUs to evaluate their performance in the SE domain, there exist other NLUs which are not included in our study. Since our goal is to find the best performing NLU in the SE domain, in our study we only select the NLUs that have been widely used by researchers and practitioners and they can be trained using their API calls and/or user interface.

Our study may be impacted by the fact that we evaluate the NLUs using the Repository and Stack Overflow tasks; hence our results may not generalize to other tasks in the SE domain. However, we believe that they cover very common tasks in SE, which could be improved with chatbots. That said, we encourage other researchers to conduct more similar studies that consider other NLUs and more SE tasks.

\section{Conclusion \& Future Work}
\label{sec:conclusions}

Software chatbots are becoming popular in the SE community due to their benefits in saving development time and resources. An NLU lies at the heart of each chatbot to enable the understanding of the user's input. Selecting the best NLU to use for a chatbot that operates in the SE domain is a challenging task. In this paper, we evaluate the performance of four widely-used NLUs, namely IBM Watson, Google Dialogflow, Rasa, and Microsoft LUIS. We assess the NLUs' performance in intents classification, confidence score, and entity extraction using two different tasks designed from a Repository and Stack Overflow contexts.

When considering all three aspects (intents classification, confidence scores, and entity extraction), we find that IBM Watson is the best performing NLU for the studied SE tasks. For each individual aspect, in intents classification, IBM Watson outperforms other NLUs for both tasks. On the other hand, when it comes to confidence scores, all NLUs (except for Dialogflow) return high confidence scores for the correctly classified intents. Also, we find that Rasa is the most trustworthy NLU in terms of confidence score. Finally, LUIS and IBM Watson achieve the best results in extracting entities from the Repository and Stack Overflow tasks, respectively.
Moreover, our results shed light on the characteristics that affect the NLUs' performance in intents classification (e.g., \# Training Samples) and entity extraction (e.g., unique entities). Therefore, we encourage researchers to develop techniques and methods that enhance the NLUs' performance for tasks with different characteristics. We believe that our work guides chatbot practitioners to select the NLU that best fits the SE task performed by their chatbots.

Our study paves the way for future work in this area. First, our results show that NLUs tend to perform well when they are trained on more examples. Therefore, we plan to examine different dataset augmentation techniques to generate more training examples for the NLUs to enhance their performance.
Also, we believe that there is a need for more studies that compare different NLUs using more datasets to benchmark NLUs in the SE context. We contribute towards this effort by making our dataset publicly available~\cite{NLUsCompScript_Link}.

\ifCLASSOPTIONcaptionsoff
\newpage
\fi

\bibliographystyle{IEEEtranN}

\bibliography{ms}
\begin{IEEEbiography}[{\includegraphics[width=1in,height=1.25in,clip,keepaspectratio]{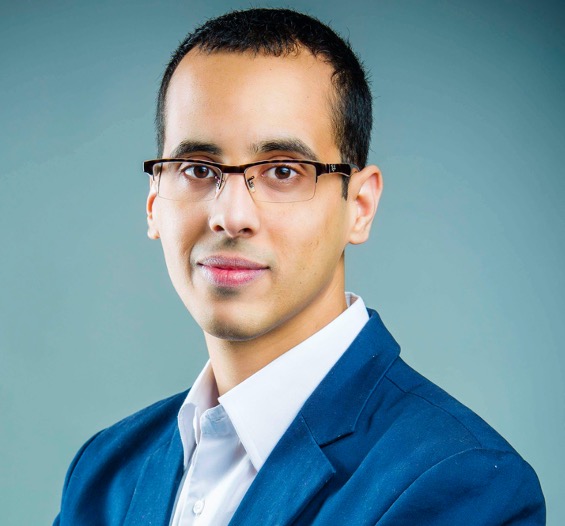}}]{Ahmad Abdellatif}is a PhD candidate in the Department of Computer Science and Software Engineering at Concordia University, Montreal, Canada. He received his masters in Software Engineering from King Fahd University of Petroleum and Minerals in 2015, where his work focused on assessing software quality. He did his Bachelors in Computer Engineering at An-Najah National University. His research interests are in software chatbots, mining software repositories, software quality, and empirical software engineering. More about his work is available at \url{http://das.encs.concordia.ca/members/ahmad-abdellatif}.
\end{IEEEbiography}

\begin{IEEEbiography}[{\includegraphics[width=1in,height=1.25in,clip,keepaspectratio]{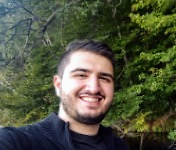}}]{Khaled Badran}is a Master's Software Engineering student at Concordia University, Montreal, Canada. He works as a research assistant in the Data-driven Analysis of Software (DAS) lab. His research interests are in software bots/chatbots, mining software repositories, natural language understanding, and empirical software engineering. He was awarded the prestigious NSERC Undergraduate Research Award. You can find more at \url{http://das.encs.concordia.ca/members/khaled-badran}.
\end{IEEEbiography}

\begin{IEEEbiography}[{\includegraphics[width=1in,height=1.25in,clip,keepaspectratio]{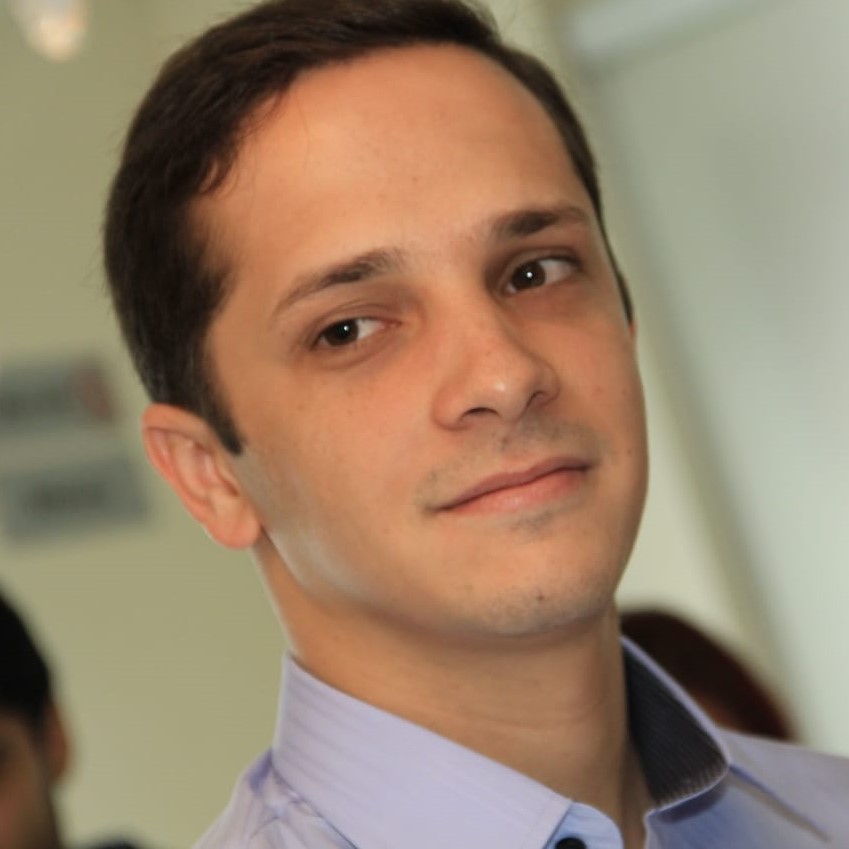}}]{Diego Elias Costa}is a postdoctoral researcher in the Department of Computer Science and Software Engineering at Concordia University. He received his PhD in Computer Science from Heidelberg University, Germany.
His research interests cover a wide range of software engineering and performance engineering related topics, including mining software repositories, empirical software engineering, performance testing, software ecosystems, and software bots. You can find more about him \url{http://das.encs.concordia.ca/members/diego-costa}.
\end{IEEEbiography}

\begin{IEEEbiography}[{\includegraphics[width=1in,height=1.25in,clip,keepaspectratio]{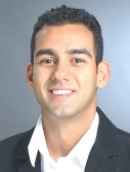}}]{Emad Shihab}is an associate professor and Concordia research chair in the Department of Computer Science and Software Engineering at Concordia University. Dr. Shihab leads the Data-driven Analysis of Software (DAS) lab. He received his PhD from Queens University. Dr. Shihab's research interests are in Software Quality Assurance, Mining Software Repositories, Technical Debt, Mobile Applications and Software Architecture. He worked as a software research intern at Research In Motion in Waterloo, Ontario and Microsoft Research in Redmond, Washington. Dr. Shihab is a member of the IEEE and ACM. More information can be found at \url{http://das.encs.concordia.ca}.
\end{IEEEbiography}
\balance

\newpage
\onecolumn
\appendix
\section{Appendix}
\label{sec:appendix}
%	\centering
%	\includegraphics[width=0.50\linewidth]{Figures/ChatbotIntentsRecall.pdf}
%	\captionof{figure}{fffffff}
%	\label{fig:ChatbotIntentsClassificationResultsRecallaa}

% To reset the numbering of the figures in the Appendix
%\setcounter{figure}{0}
\noindent\textbf{Detailed Precision and Recall Values When Evaluating Intents Classifications.}\\
In section~\ref{subsec:intents_classification_results}, we presented the F1-measure values for each task when classifying intents. In this appendix, we add detailed precision, recall, and F1-measure values that are used to compute the F1-measure values for each task on each NLU.
\begin{table*}[htp]
	\caption{Intents classification results for the Repository task.}
	\label{table:chatbot_all_intents_results}
	\centering
%	\hspace*{-2.2cm}
%	\scalebox{0.87}{
\makebox[\linewidth]{
		\begin{tabular}{@{}l|ccc|ccc|ccc|ccc@{}}
			\toprule
			\multicolumn{1}{c|}{\multirow{2}{*}{\textbf{Intent}}} & \multicolumn{3}{c|}{\textbf{IBM Watson}} & \multicolumn{3}{c|}{\textbf{Dialogflow}} & \multicolumn{3}{c|}{\textbf{Rasa}} & \multicolumn{3}{c}{\textbf{LUIS}} \\ \cmidrule(l){2-13} 
			\multicolumn{1}{c|}{} & \textbf{P} & \textbf{R} & \textbf{F1} & \textbf{P} & \textbf{R} & \textbf{F1} & \textbf{P} & \textbf{R} & \textbf{F1} & \textbf{P} & \textbf{R} & \textbf{F1} \\ \midrule
			\textbf{\CountCommitsByDates} & 86.4 & 90.5 & 88.4 & 84.0 & 100.0 & 91.3 & 100.0 & 66.7 & 80.0 & 88.9 & 38.1 & 53.3 \\
			\textbf{\FileCommits} & 83.3 & 41.7 & 55.6 & 100.0 & 66.7 & 80.0 & 100.0 & 16.7 & 28.6 & 16.7 & 8.3 & 11.1 \\
			\textbf{\OverloadedDev} & 100.0 & 88.9 & 94.1 & 100.0 & 66.7 & 80.0 & 88.9 & 88.9 & 88.9 & 40.9 & 100.0 & 58.1 \\
			\textbf{\CommitsByDate} & 87.5 & 58.3 & 70.0 & 50.0 & 50.0 & 50.0 & 88.9 & 66.7 & 76.2 & 0.0 & 0.0 & 0.0 \\
			\textbf{\ExperiencedDevFixBugs} & 100.0 & 85.7 & 92.3 & 92.3 & 85.7 & 88.9 & 100.0 & 85.7 & 92.3 & 100.0 & 7.1 & 13.3 \\
			\textbf{\BuggyFiles} & 100.0 & 92.3 & 96.0 & 100.0 & 92.3 & 96.0 & 100.0 & 100.0 & 100.0 & 90.9 & 76.9 & 83.3 \\
			\textbf{\BuggyCommitsByDate} & 60.0 & 92.3 & 72.7 & 80.0 & 61.5 & 69.6 & 70.6 & 92.3 & 80.0 & 100.0 & 84.6 & 91.7 \\
			\textbf{\FixCommit} & 100.0 & 100.0 & 100.0 & 84.6 & 100.0 & 91.7 & 100.0 & 100.0 & 100.0 & 52.4 & 100.0 & 68.8 \\
			\textbf{\BuggyCommit} & 90.0 & 100.0 & 94.7 & 90.0 & 100.0 & 94.7 & 90.0 & 100.0 & 94.7 & 47.4 & 100.0 & 64.3 \\
			\textbf{\BuggyFixCommits} & 100.0 & 100.0 & 100.0 & 85.7 & 85.7 & 85.7 & 50.0 & 85.7 & 63.2 & 33.3 & 85.7 & 48.0 \\ \bottomrule
		\end{tabular}}
%	}
	\begin{tablenotes}
%		\addtolength{\itemindent}{-0.4cm}
		\addtolength{\itemindent}{1.4cm}
		\small
		\item P: Precision, R: Recall, F1: F1-measure
	\end{tablenotes}
\end{table*}

\begin{table*}[htp]
	\caption{Intents classification results for the Stack Overflow task.}
	\label{table:sof_all_intents_results}
	\centering
%	\scalebox{0.87}{
\makebox[\linewidth]{
		\begin{tabular}{@{}l|ccc|ccc|ccc|ccc@{}}
			\toprule
			\multicolumn{1}{c|}{\multirow{2}{*}{\textbf{Intent}}} & \multicolumn{3}{c|}{\textbf{IBM Watson}} & \multicolumn{3}{c|}{\textbf{Dialogflow}} & \multicolumn{3}{c|}{\textbf{Rasa}} & \multicolumn{3}{c}{\textbf{LUIS}} \\ \cmidrule(l){2-13} 
			\multicolumn{1}{c|}{} & \textbf{P} & \textbf{R} & \textbf{F1} & \textbf{P} & \textbf{R} & \textbf{F1} & \textbf{P} & \textbf{R} & \textbf{F1} & \textbf{P} & \textbf{R} & \textbf{F1} \\ \midrule
			\textbf{\UsingMethodImproperly} & 93.6 & 73.3 & 79.1 & 84.0 & 58.3 & 66.5 & 81.6 & 58.3 & 64.4 & 71.7 & 51.7 & 57.5 \\
			\textbf{\LookingForCodeSample} & 85.3 & 97.9 & 90.9 & 81.5 & 91.4 & 85.9 & 78.2 & 92.1 & 84.5 & 75.4 & 95.7 & 84.0 \\
			\textbf{\FacingError} & 80.0 & 80.0 & 80.0 & 60.0 & 60.0 & 60.0 & 30.0 & 40.0 & 33.3 & 10.0 & 10.0 & 10.0 \\
			\textbf{\PassingData} & 35.0 & 40.0 & 36.7 & 50.0 & 50.0 & 50.0 & 35.0 & 40.0 & 36.7 & 0.0 & 0.0 & 0.0 \\
			\textbf{\LookingForBestPractice} & 86.7 & 80.0 & 81.3 & 86.7 & 80.0 & 81.3 & 76.7 & 80.0 & 78.0 & 90.0 & 80.0 & 83.3 \\ \bottomrule
		\end{tabular}}
%	}
	\begin{tablenotes}
				\addtolength{\itemindent}{2cm}
		\small
		\item P: Precision, R: Recall, F1: F1-measure
	\end{tablenotes}
\end{table*}

%\newpage

\noindent\textbf{Detailed Precision and Recall Values When Assessing Entity Extraction.}\\
When discussing the entity extraction results in Section~\ref{subsec:entityExtractionResult}, we presented the F1-measure values for each task. Here, we present the detailed precision, recall, and F1-measure values that are used to compute the F1-measure values presented earlier.

%\begin{figure*}[htp]	
%	\centering		
%	\begin{subfigure}{0.5\textwidth}	
%		\centering	
%		\includegraphics[width=1.00\linewidth]{Figures/ChatbotEntitiesPrecision.pdf}	
%		\caption{Precision}
%	\end{subfigure}\hfill	
%	\begin{subfigure}{0.5\textwidth}	
%		\centering	
%		\includegraphics[width=1.00\linewidth]{Figures/ChatbotEntitiesRecall.pdf}
%		\caption{Recall}	
%	\end{subfigure}\hfill	
%	\caption{The distribution of the entity extraction precision and recall per entity type in the Repository task.}	
%	\label{fig:ChatbotEntitiesExtractionResults}	
%\end{figure*}
%
%\begin{figure*}[htp]	
%	\centering		
%	\begin{subfigure}{0.5\textwidth}	
%		\centering	
%		\includegraphics[width=1.00\linewidth]{Figures/SOFEntitiesPrecision.pdf}	
%		\caption{Precision}
%	\end{subfigure}\hfill	
%	\begin{subfigure}{0.5\textwidth}	
%		\centering	
%		\includegraphics[width=1.00\linewidth]{Figures/SOFEntitiesRecall.pdf}
%		\caption{Recall}	
%	\end{subfigure}\hfill	
%	\caption{The distribution of the entity extraction precision and recall per entity type in the Stack Overflow task.}	
%	\vspace{-2in}
%	\label{fig:SOFEntitiesExtractionResults}	
%\end{figure*}

\begin{table*}[htp]
	\caption{Entity extraction classification results for the Repository task.}
	\label{table:chatbot_all_entities_results}
	\centering

%	\scalebox{0.87}{
\makebox[\linewidth]{
		\begin{tabular}{@{}l|ccc|ccc|ccc|ccc@{}}
			\toprule
			\multirow{2}{*}{\textbf{Entity Type}} & \multicolumn{3}{c|}{\textbf{IBM Watson}} & \multicolumn{3}{c|}{\textbf{Dialogflow}} & \multicolumn{3}{c|}{\textbf{Rasa}} & \multicolumn{3}{c}{\textbf{LUIS}} \\ \cmidrule(l){2-13} 
			& \textbf{P} & \textbf{R} & \textbf{F1} & \textbf{P} & \textbf{R} & \textbf{F1} & \textbf{P} & \textbf{R} & \textbf{F1} & \textbf{P} & \textbf{R} & \textbf{F1} \\ \midrule
			\textbf{\FileName} & 8.6 & 100.0 & 15.9 & 86.4 & 73.1 & 79.2 & 93.8 & 57.7 & 71.4 & 83.3 & 76.9 & 80.0 \\
			\textbf{\JiraTicket} & 100.0 & 100.0 & 100.0 & 84.6 & 100.0 & 91.7 & 100.0 & 81.8 & 90.0 & 100.0 & 100.0 & 100.0 \\
			\textbf{\DateTime} & 78.1 & 96.2 & 86.2 & 56.9 & 55.8 & 56.3 & 100.0 & 100.0 & 100.0 & 98.1 & 98.1 & 98.1 \\
			\textbf{\CommitHash} & 100.0 & 100.0 & 100.0 & 100.0 & 100.0 & 100.0 & 100.0 & 80.0 & 88.9 & 100.0 & 100.0 & 100.0 \\
%			\textbf{No Entity} & 100.0 & 95.5 & 97.7 & 88.0 & 100.0 & 93.6 & 58.3 & 95.5 & 72.4 & 91.7 & 100.0 & 95.7  \\ 
			\bottomrule
		\end{tabular}}
%	}
	\begin{tablenotes}
		\addtolength{\itemindent}{1.9cm}
		\small
		\item P: Precision, R: Recall, F1: F1-measure
%		\vspace{-2in}
	\end{tablenotes}

\end{table*}

\begin{table*}[htp]
	\caption{Entity extraction classification results for the Stack Overflow task.}
	\label{table:sof_all_entities_results}
	\centering
%	\scalebox{0.87}{
\makebox[\linewidth]{
		\begin{tabular}{@{}l|ccc|ccc|ccc|ccc@{}}
			\toprule
			\multirow{2}{*}{\textbf{Entity Type}} & \multicolumn{3}{c|}{\textbf{IBM Watson}} & \multicolumn{3}{c|}{\textbf{Dialogflow}} & \multicolumn{3}{c|}{\textbf{Rasa}} & \multicolumn{3}{c}{\textbf{LUIS}} \\ \cmidrule(l){2-13} 
			& \textbf{P} & \textbf{R} & \textbf{F1} & \textbf{P} & \textbf{R} & \textbf{F1} & \textbf{P} & \textbf{R} & \textbf{F1} & \textbf{P} & \textbf{R} & \textbf{F1} \\ \midrule
			\textbf{\ProgLanguage} & 88.4 & 96.2 & 92.0 & 90.7 & 97.2 & 93.7 & 92.1 & 91.5 & 91.4 & 95.9 & 80.2 & 86.8 \\
			\textbf{\Framework} & 70.7 & 63.2 & 65.4 & 85.3 & 43.2 & 56.0 & 89.1 & 42.1 & 56.1 & 88.6 & 41.1 & 54.4 \\
			\textbf{\Standards} & 81.0 & 42.9 & 54.1 & 85.7 & 47.6 & 56.9 & 17.5 & 14.3 & 14.9 & 23.8 & 14.3 & 17.5 \\
			\textbf{\API} & 50.9 & 39.2 & 43.3 & 84.0 & 32.4 & 42.8 & 62.3 & 20.3 & 29.9 & 48.6 & 16.2 & 23.5 \\
			\textbf{\Platform} & 76.9 & 61.5 & 67.4 & 64.1 & 53.8 & 55.9 & 84.6 & 69.2 & 75.1 & 53.8 & 38.5 & 43.6 \\
%			\textbf{No Entity} & 76.7 & 76.9 & 74.7 & 45.9 & 94.9 & 61.0 & 43.0 & 97.4 & 59.1 & 36.4 & 97.4 & 52.5 \\ 
			\bottomrule
		\end{tabular}}
%	}
\begin{tablenotes}
			\addtolength{\itemindent}{2.7cm}
	\small
	\item P: Precision, R: Recall, F1: F1-measure
\end{tablenotes}
\end{table*}

\clearpage
\end{document}